\documentclass[fleqn,usenatbib]{mnras}

\usepackage[T1]{fontenc}
\usepackage{ae,aecompl}

\usepackage{graphicx}	
\usepackage{amsmath}	
\usepackage{amssymb}	
\usepackage{gensymb}
\usepackage[dvipsnames]{xcolor}

\usepackage{ulem}
 \usepackage{txfonts}

\title[99 Herculis]{Formation of the polar debris disc around 99 Herculis}

\author[Smallwood et al.]{Jeremy L. Smallwood,$^{1}$\thanks{E-mail: smallj2@unlv.nevada.edu}
Alessia Franchini,$^{1}$
Cheng Chen,$^{1}$
Eric Becerril,$^{1}$
\newauthor
Stephen H. Lubow,$^{2}$
Chao-Chin Yang$,^{1}$
and Rebecca G. Martin$^{1}$
\\
$^1$Department of Physics and Astronomy, University of Nevada, Las Vegas, 4505 South Maryland Parkway, Las Vegas, NV 89154, USA\\
$^2$Space Telescope Science Institute, Baltimore, MD 21218, USA
}

\date{Accepted XXX. Received YYY; in original form ZZZ}

\pubyear{2019}

\begin{document}
\label{firstpage}
\pagerange{\pageref{firstpage}--\pageref{lastpage}}
\maketitle

\begin{abstract}
  We investigate the formation mechanism for the observed nearly polar aligned  (perpendicular to the binary orbital plane) debris ring around the eccentric orbit binary 99 Herculis. 
 An initially inclined nonpolar debris ring or disc will not remain flat and will not evolve to a polar configuration, due to the effects of differential nodal precession that alter its flat structure. However, a gas disc with embedded well coupled solids around the eccentric binary may evolve to a polar configuration as a result of pressure forces that maintain the disc flatness and as a result of viscous dissipation that allows the disc to increase its tilt. Once the gas disc disperses, the debris disc is in a polar aligned state in which there is little precession.
 We use three-dimensional hydrodynamical simulations, linear theory, and particle dynamics to study the evolution of a misaligned circumbinary gas disc and explore the effects of the initial disc tilt, mass, and size. We find that for a wide range of parameter space, the polar alignment timescale is shorter than the lifetime of the gas disc. 
 Using the observed level of alignment  of $3\degree$  from polar, we place an upper limit on the mass of the gas disc of about $0.014 \,M_\odot$ at the time of gas dispersal. We conclude that  the polar debris disc around 99 Her can be explained as the result of an initially moderately inclined gas disc with embedded solids.  Such a disc may provide an environment for the formation of polar planets.
\end{abstract}




\section{Introduction}





The majority of stars that form within dense regions of stellar clusters are formed as binary systems \citep{Duquennoy1991,Ghez1993,Duchene2013}, which are most likely accompanied by circumstellar and circumbinary discs \cite[e.g.,][]{Dutrey1994,Beust2005}. 
The evolution of circumbinary disc structure and orientation has been studied extensively. 
An initially slightly misaligned misaligned  circumstellar or circumbinary disc involving a circular orbit binary precesses about the binary angular momentum vector and evolves towards alignment with it due to viscous dissipation in the disc.  As a  result, the disc becomes coplanar  with the binary \citep{papaloizou1995,Lubow2000,Nixon2011,Facchini2013,Foucart2014}.
If the binary orbit is eccentric, a low mass circumbinary disc with a large enough initial inclination can precess around the eccentricity vector   (semi-major axis) of the binary.  The disc's angular momentum vector eventually aligns with the eccentricity vector. This means that the disc angular momentum is aligned polar (perpendicular) with respect to the binary angular momentum \citep{Aly2015,Martinlubow2017,Lubow2018,Zanazzi2018,Martin2018}. The disc then lies perpendicular to the orbital plane of the binary.  A massive disc aligns to a generalised polar state at lower misalignment to the binary orbital plane \citep{Zanazzi2018, MartinLubow2019}.

Disc misalignment can occur various phases of stellar evolution.
Misaligned discs around binaries can arise from chaotic accretion of turbulent molecular clouds in star-forming regions \citep{Offner2010,Tokuda2014,Bate2012} or whenever a young binary system accretes material post-formation \citep{Bate2010,Bate2018}. Furthermore, misalignments can be present if the binary forms within an elongated cloud, where the binary axis is misaligned with respect to the cloud rotation axis \cite[e.g.,][]{Bonnell1992}.  

There are currently a number of observed systems with misaligned circumbinary discs. KH 15D is an eccentric spectroscopic binary T Tauri star with a misaligned circumbinary disc \citep{Chiang2004,Winn2004,Lodato2013,Smallwood2019}. High misalignment has been observed in the binary protostar IRS 43, where the tilt of the disc is at least $60\degree$ with respect to the orbital plane of the binary \citep{Brinch2016}. The binary GG Tau consists of T Tauri stars with a circumbinary disc misaligned by $25\degree$--$30\degree$ along with misaligned discs around each of the binary components \citep{Dutrey1994,Kohler2011,Cazzoletti2017,Aly2018}. The binary system HD 98800 BaBb shows evidence of having a nearly polar circumbinary gas disc \citep{Kennedy2019}. Lastly, misalignment can be observed also after the gas disc has been dispersed. The binary 99 Herculis (99 Her) has a misaligned circumbinary debris disc that is almost perpendicular to the binary orbital plane \citep{Kennedy2012}.



 The lifetimes of discs around single stars are observed to be around $1$--$10\, \rm Myr$ \citep{Haisch2001,Hernandez2007,Hernandez2008,Mamajek2009,Ribas2015}. Mass accretion rates through circumbinary discs may be inhibited due to the tidal torques exerted by the binary, resulting in extended disc lifetimes \cite[e.g.,][]{Alexander2012}.  There is observational evidence for extended disc lifetimes for circumbinary discs. For example, the circumbinary gas discs HD 98800 B, V4046 Sgr, and AK Sco  have disc ages of $10\pm 3 \, \rm Myr$, $23\pm 3\, \rm Myr$, and $18\pm 1\, \rm Myr$, respectively \citep{Soderblom1998,Mamajek2014,Czekala2015}. 
 
 The lifetime of protoplanetary disks is fundamentally linked to their dispersal mechanisms. The main processes that remove mass and/or angular momentum from the disc include viscous evolution of the disc \cite[e.g.,][]{shakura1973,Lynden-Bell1974,Hartmann1998}, photoevaporation by stellar radiation \cite[e.g.,][]{Shu1993,Hollenbach1994}, magnetically-launched jets and winds \cite[e.g.,][]{Konigl2011}, and the interaction with newborn planets \cite[e.g.,][]{Kley2012}. However, the $\sim \rm Myr$ timescale for planet formation \citep{Pollack1996} suggests that this is not a major factor for rapid disc dispersal required by observations \citep{Alexander2014}. It has also been shown that planets likely account for $\lesssim 1\%$ of the initial disc mass budget \citep{Wright2011,Mayor2011}. 
 
 After the gaseous protoplanetary disc is dispersed, the remnant planetesimals produce a second generation of dust through collisions which leads to the formation of a gas-poor, less massive disc called a debris disc. These debris discs are much cooler in temperature and are analogous to the Solar system Kuiper belt \cite[e.g.,][]{Wyatt2008,Hughes2018}. 
 

The larger solids in a debris disc can be modeled as set of particles on  nearly circular ballistic circumbinary orbits. The particles are not interacting except during close encounters
or collisions. If such a disc is initially inclined with respect to the binary by some arbitrary tilt angle, the orbits will undergo differential nodal
precession. As a consequence, the disc will not maintain its flat form and initially nearby orbits
may undergo violent collisions \citep[e.g.,][]{Nesvold2016}. A low mass polar (or coplanar)  debris disc is an exception to this
rule because it does not undergo nodal precession. 

\cite{Kennedy2012} suggested that the polar aligned debris disc in 99 Her could be the result of the capture of material or a stellar exchange  which leaves the circumbinary debris in a polar orbit. Although such a scenario is possible, it involves fine tuning of the conditions that result in a polar state, as they point out. \cite{Martinlubow2017}  instead suggested that the 99 Her debris disc began as a somewhat inclined gas disc with embedded solids that later evolved to a polar configuration. Since gas discs can evolve to a polar configuration, small embedded solids in the disc should follow the gas to a polar configuration. During this tilt evolution, dust (and perhaps somewhat larger solid bodies) can be well coupled dynamically to the gas that evolves as a nearly flat disc due to gas pressure communication. The disc tilt evolves as a consequence of viscous dissipation in the gas. Once in the polar configuration, the disc evolved to become a debris disc in the usual manner. This process might then operate over a wide range of initial conditions without fine tuning. The purpose of this paper is to explore the viability of this scenario.

Circumbinary dust particles experience various degrees of coupling to the gas depending on their Stokes number \cite[e.g.,][]{Birnstiel2010}. Over time, small, initially well coupled, dust grains grow to higher Stokes number and gradually decouple from the gas disc. If significant decoupling occurs in a misaligned circumbinary disc,  the dust particle orbits may evolve  independently of the gas disc  \citep[e.g.,][]{Aly2020}.  Thus the gas disc around 99 Her must evolve to a polar configuration before the dust decouples because a polar debris ring is observed in 99 Her.
According to the dust coagulation model, dust grains remain coupled to the gas with $\rm St < 1$ \citep{Birnstiel2012}. Large decimetre-sized dust grains have large relative velocities which induce fragmentation rather than sticking during collisions. This fragmentation barrier limits dust growth which in-turn prevents the dust from decoupling from the gas \citep{Brauer2008,Blum2008,Birnstiel2010}. Nevertheless, planet formation may still proceed when the dust particles are concentrated to high density and gravitationally collapse into planetesimals, e.g., by the streaming instability \cite[e.g.,][]{Johansen2012,Yang2017}. Therefore, if the fragmentation barrier dominates in the disc of 99 Her, a significant fraction of the solid materials may still remain coupled to the gas while the gas disc evolves to a polar state.



The structure of this paper is as follows. In Section \ref{sec:99Her} we discuss the orbital parameters of 99 Her and the observational evidence for a polar aligned debris disc. In Section \ref{sec:part_orbits}, we show the orbital evolution of a circumbinary particle in the potential of the binary system 99 Her. In Section \ref{sec:SPHsims}, we present the setup for the SPH simulations and discuss the results. In Section \ref{sec::linear_theory}, we apply linear theory for the disc evolution. Finally, we discuss the possible implications of this work in Section \ref{sec:dis} and  we draw  our conclusions in Section \ref{sec:conc}.

\section{Observed properties of 99 Herculis}
\label{sec:99Her}

99 Her is a particularly interesting system since it hosts a resolved polar debris disc and the binary orbit is well characterized. The binary has been observed since the latter half of the 1800's \cite[e.g.,][]{Burnham1878,Flammarion1879,Gore1890}. It consists of an F7V primary and a K4V secondary. The primary star has an estimated age of $6$--$10$ Gyr, consistent with being on the main-sequence \citep{Nordstrom2004,Takeda2007}. 
There has been an abundance of observations of this system in recent years.  \cite{Kennedy2012} performed a more precise derivation of the system orbital parameters. The orbit was found by fitting position angles (PAs) and separations from the Washington Double Star (WDS) Catalog \citep{Mason2018}. The semimajor axis, eccentricity, inclination and orbital period  are $a=16.5\, \rm au$, $e_{\rm b}=0.766$, $i=39\degree$ and $P_{\rm orb}=56.3\, \rm yr$, respectively.  The longitude of the ascending node and longitude of pericentre are $\Omega=41 \degree$ and $\omega=116\degree$ respectively.  The longitude of pericentre is measured anticlockwise from the ascending node, and projected on to the sky plane which has a position angle of $163\degree$. The reason that the sum of $\Omega$ and $\omega$ do not equal the sky plane position angle is due to the fact that the binary orbit is inclined. The total mass of the binary inferred from the data is $M=1.4 \, \rm M_{\odot}$ using a distance of $15.64\, \rm pc$ \citep{vanLeeuwen2008}. 
Using the spectroscopic mass function, a mass ratio of $0.49$ was inferred. Therefore the primary mass is $M_1=0.94\, \rm M_{\odot}$ while the mass of the secondary is $M_2=0.46 \, \rm M_{\odot}$.

Debris discs around binary systems are as common as around single star systems \citep{Trilling2007}. The  debris around 99 Her  was first detected using $100$ and $160\, \rm \mu m$ data from {\it Herschel} Photodetector and Array Camera and Spectrometer \cite[PACS;][]{Poglitsch2010,Griffin2010}. The debris was first detected but not resolved by SPIRE at $250$ and $350\, \rm \mu m$. \cite{Kennedy2012} provided resolved PACS images of the disc and were able to estimate the debris structure, inclination and position angle using two-dimensional Gaussian models. To best fit the observations, they invoked a debris ring model located  in a small range of radii near $120 \, \rm au$. The grain properties and size distributions cannot accurately be constrained due to the low number of measured disc emission wavelengths.

The projection of the binary pericentre direction on the plane of the sky has a PA of $163\degree \pm 2\degree$, and a line perpendicular to this has a PA of $73\degree \pm 2\degree$. Since the observed debris disc has a PA of $72\degree$, \cite{Kennedy2012} concluded that this is consistent with the disc being at $87\degree$ with respect to the binary pericentre direction, or $3^\circ$ away from polar alignment.  However, they found that the ring could be mirrored in the sky plane causing the observed misalignment to be $30\degree$.

\citet{Kennedy2012} used circumbinary test particles around the eccentric binary 99 Her to model each of these possible ring misalignments.  The test particles undergo secular perturbations which lead to nodal precession, as well as inclination oscillations. A debris ring with a tilt of  $30\degree$ will eventually spread out into a broader, non-ring like structure due  to the effects of differential nodal precession. By assuming the largest fragments are at least $1\, \rm mm$, the secular precession period is estimated to be $0.5\, \rm Myr$ \citep{Kennedy2012}. By their calculations, within $10$ precession  periods ($5\, \rm Myr$), the ring structure will be erased. Thus, the model that best fits the PACS images with considerable agreement is a ring in a nearly polar configuration. 

The polar particle orbits in their model are stable over the stellar lifetime, which means that the observed dust could be the steady-state collision products of the polar planetesimal ring. Though a thin ring of debris best fit their data, the initial extent of a gas disc must have been at least as large as this. Radial drift can reduce the outer radius of the debris disc due to the gas drag force \citep{Adachi1976,Weidenschilling1977,Takeuchi2002}. This process can lead to the outer radius of the dust being less than the outer radius of the gas.

\begin{figure}
\centering
\includegraphics[width=\columnwidth]{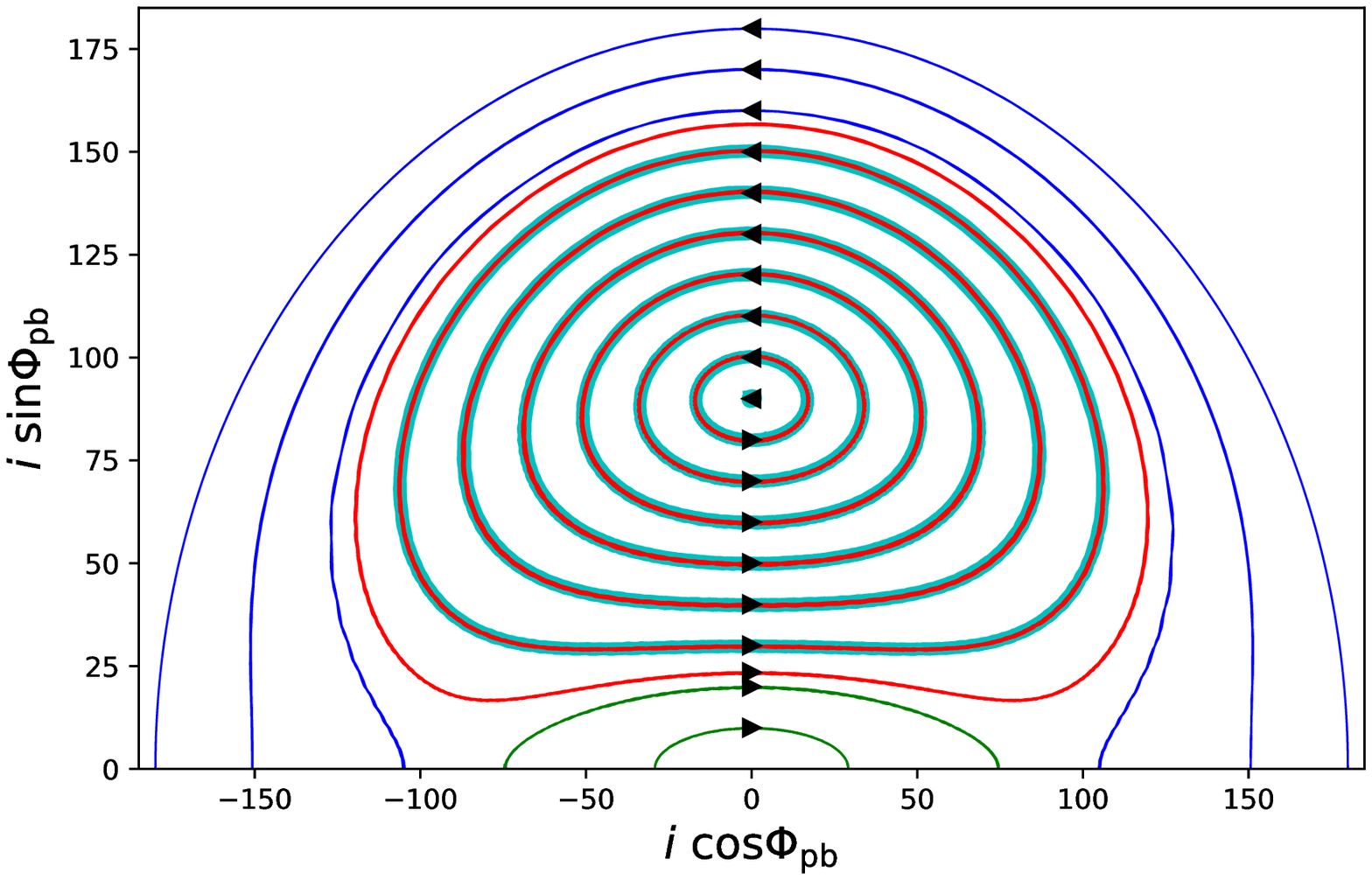}
\includegraphics[width=\columnwidth]{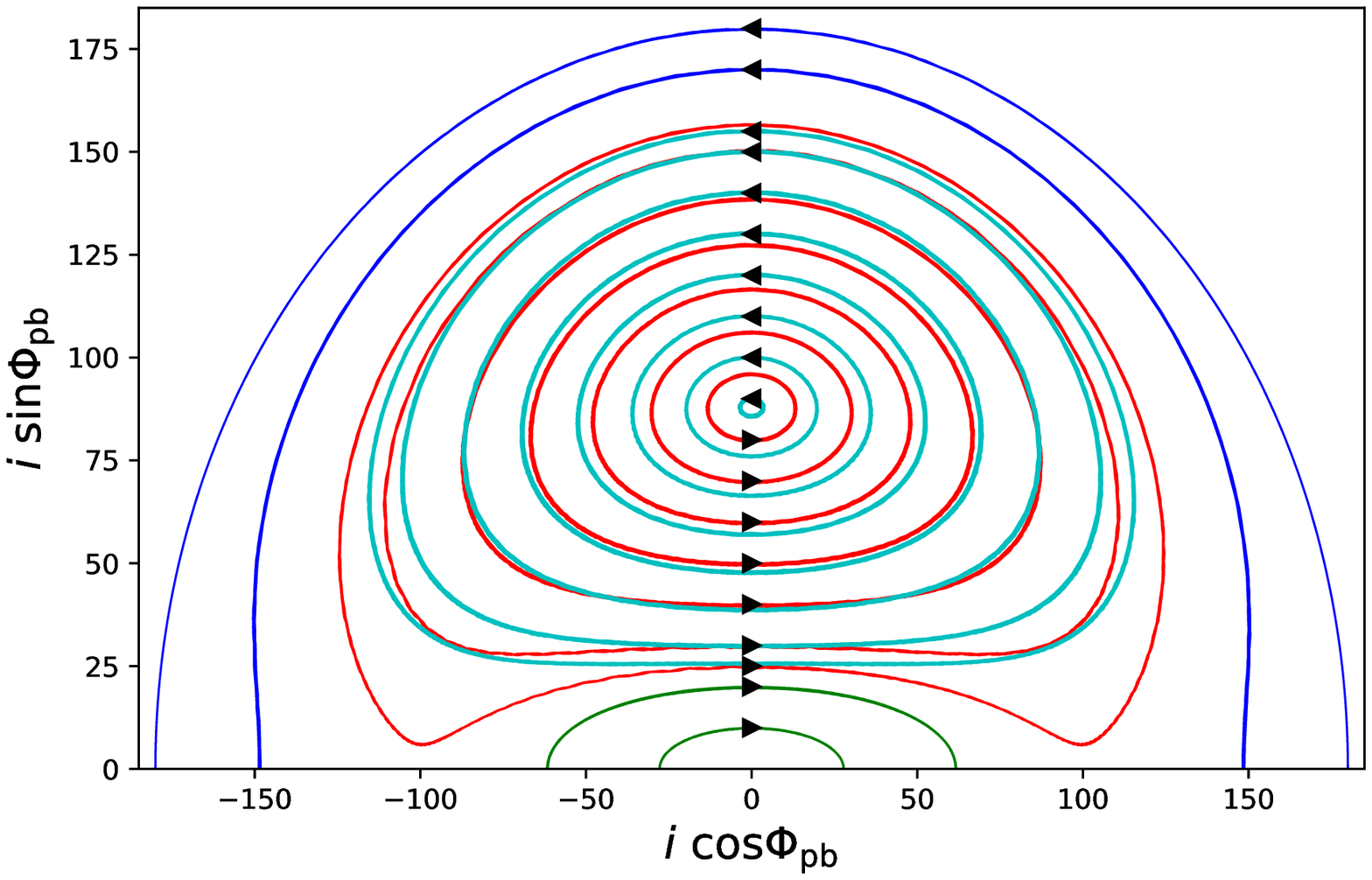}
\includegraphics[width=\columnwidth]{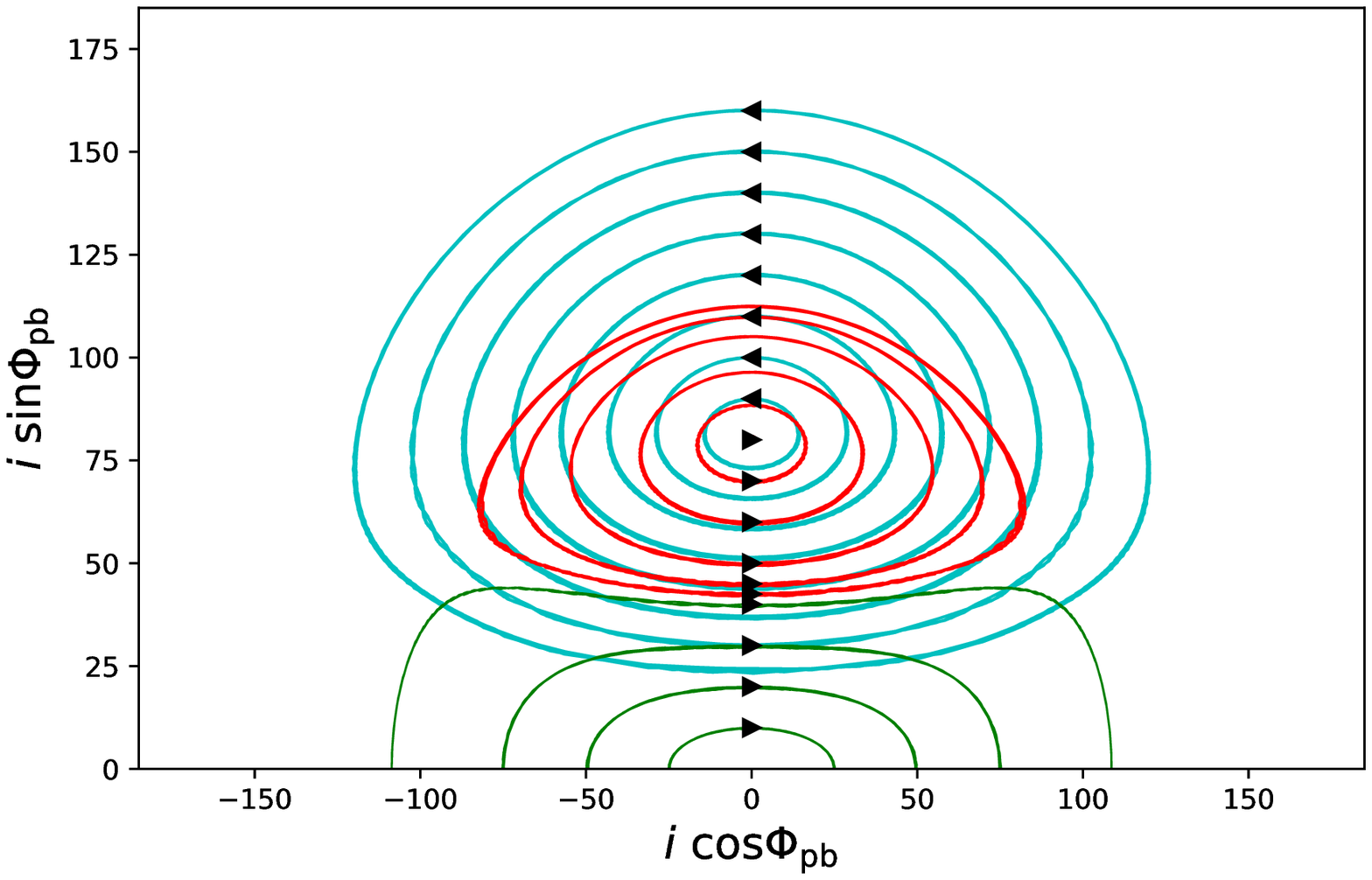}
\caption{The $i\cos \phi_{\rm pb}$ -- $i \sin \phi_{\rm pb}$ phase portrait for an initially circular particle orbits around the 99 Her binary system for different initial inclination, $i$, and initial  precession angle, $\phi_{\rm pb}=\phi_0 = 90\degree$. The initial separation of the test particle is $d = 82.5\, \rm au$ in all panels.  The green lines show  prograde circulating solutions. The red lines show librating solutions that have initial inclination $i<i_{\rm s}$, the stationary inclination. The cyan lines show librating solutions that have initial inclination $i>i_{\rm s}$  and the blue lines show the retrograde circulating solutions. The black arrows represent the initial position of the particle in the phase portrait and its orbital direction.
Upper panel: test particle orbits with $j=0.0$. 
Middle panel: particle orbits with $j=0.1$. 
Lower panel: particle orbits with $j=0.51$.  
}
\label{fig:test_part}
\end{figure}

\section{Inclined circumbinary particle orbits around 99 Her}
\label{sec:part_orbits}


We first investigate the evolution of inclined circumbinary particle orbits in the potential of the binary system 99 Her, following \cite{Chen2019}.  For a circular orbit binary, the orbital angular momentum of a circumbinary test (massless) particle always precesses about the binary angular momentum vector. We adopt a Cartesian coordinate system $(x,y,z)$, where the $x$--axis is along the initial binary eccentricity vector and the $z$--axis is along the initial binary angular momentum vector.
The origin of the coordinate system is taken to be the initial center of mass of the entire system.
An eccentric orbit binary generates a secular potential that is nonaxisymmetric about the $z$--axis.
For test particle simulations, the binary orbit is fixed. However, for a massive particle the binary eccentricity vector precesses. In addition, the particle angular momentum vector oscillates. 
 
We consider nearly circular circumbinary particle orbits with an initial semi--major axis of $d = 82.5\, \rm au$ (five times the binary semimajor axis) with  inclination $i_0$  with respect to the binary orbital plane.
We consider test particles with various inclinations that have stable orbits.
We define the precession angle, $\phi_{\rm pb}$, of the particle angular momentum vector relative to the binary eccentricity vector as  
 \begin{equation}
 \phi_{\rm pb} = \phi - \phi_{\rm b}.
 \end{equation}
The longitude of the ascending node of the particle relative to the $x$-axis (the initial binary eccentricity vector) is given by
 \begin{equation}
    \phi = \tan^{-1} \bigg( \frac{l_{\rm py}}{l_{\rm px}} \bigg) + \pi/2,
    \label{eq::phi_p} 
\end{equation}
where $l_{\rm px},l_{\rm py}$ are the $x$ and $y$ components of the particle angular momentum vector ${\bf l}_{\rm p}$.
  The azimuthal angle of the eccentricity vector is defined as 
\begin{equation}
    \phi_{\rm b} = \tan^{-1}  \bigg( \frac{e_{\rm by}}{e_{\rm bx}} \bigg) + \pi/2,
    \label{eq::phi_b}
\end{equation}
where $e_{\rm bx},e_{\rm by}$ are the $x$ and $y$ components of the eccentricity vector ${\bf e}_{\rm b}$. The initial precession angle of the particle is $\phi_{\rm pb}=\phi_0 = 90\degree$ in all of the three body simulations we consider.  We define the angular momentum ratio of the particle to the binary as
\begin{equation}
j=\frac{J_{\rm p}}{J_{\rm b}}.
\end{equation}
We consider three different values of $j=0$, $0.1$, and $0.51$ by varying the mass of the particle to 0, 0.00645 and 0.03409 ${\rm M_{\odot}}$, while keeping the particle separation fixed. 


\begin{table*}
	\centering
	\caption{Parameters of the initial circumbinary disc for all simulations. The first column is the model name. The second column is the Figure number in which the model appears. The third column is the initial number of particles. The fourth column is the initial disc tilt. The fifth column is the initial outer radius of the disc. The sixth column is the initial disc mass. The seventh column is the initial ratio of the angular momentum of the disc to the angular momentum of the binary. The eighth column describes whether the disc is undergoing circulation (C) or libration (L). The ninth colum in the tilt decay timescale calculated from Equation \ref{tilt_decay}. The tenth column is the approximate polar alignment timescale. The eleventh column describes whether the disc breaks. Finally, the twelfth column denotes the final disc inclination. }
	\begin{tabular}{clcccccccccc} 
		\hline
	    Model & Fig. & $N$ & Tilt  & $R_{\rm out}$ & $M_{\rm disc}$ & $j_0$ & C/L & $\tau$ & $t_{\rm polar}$& Disc Breaking & $i_{\rm final}$ \\
         &  & & $(\degree)$   &$(\rm au)$ & $(M_{\odot})$ & & & (yr) & (yr) & & $(\degree)$\\
		\hline
        \hline
		A & --  & $300,000$ & $20$  & $120$ & $0.001$ & 0.01 & C & -- & -- & No & --\\
        B & \ref{inc_plots} (upper )& $300,000$ & $30$  & $120$ & $0.001$ & 0.01 & L & $11,500$ & $49,500$ & Yes & $89$ \\
        C & \ref{inc_plots} (middle)& $300,000$ & $40$  & $120$ & $0.001$ & 0.01  &L & $11,700$ & $30,000$ &  Yes & $89$ \\
        D & \ref{inc_plots} (lower), \ref{splash}& $300,000$ & $60$  & $120$ & $0.001$ & 0.01 &L & $17,000$ & $26,000$ & No & $89$ \\
        \hline
        E & \ref{mass_plots40} &$300,000$& $40$  & $120$ & $0.01$ & 0.10 &L & $15,000$ & $36,000$ & Yes & $87$ \\
        F & -- &$300,000$ & $40$  & $120$ & $0.05$  & 0.51 & C  & -- & -- & No & --\\
        \hline
        G & \ref{mass_plots60} (upper) &$300,000$ & $60$  & $120$ & $0.01$ & 0.10 & L  & $18,000$ & $31,000$ & No & $87$ \\
        H & \ref{mass_plots60} (lower)&$300,000$ & $60$  & $120$ & $0.05$  & 0.51 &L  & $67,000$ & -- & No & $74.5$\\
        \hline
        I & \ref{fig::o200ab} (upper), \ref{fig::params_run9}  &$500,000$ & $60$  & $200$ & $0.001$ & 0.012 & L & -- & $ 67,000$ & No & $89$  \\
        J & \ref{fig::o200ab} (lower)  &$500,000$ & $60$  & $200$ & $0.01$ & 0.12 & L & -- & $45,000$ & No & $85$ \\
        \hline
	\end{tabular}
    \label{table2}
\end{table*}

The upper panel of Figure~\ref{fig:test_part} shows the circumbinary massless test particle orbits in the $i\cos \phi_{\rm pb}$--$i\sin \phi_{\rm pb}$ phase space, where $i$ is the inclination between the angular momentum of the particle and the binary.  For a massless (test) particle, we have $j=0$. For the eccentric binary orbit, the orbit of a particle with sufficiently low inclination precesses about the binary angular momentum (this is the circulating phase). However, the evolution is different to the circular orbit binary case because the particle orbit displays tilt oscillations during this process \citep{Smallwood2019}. If the test particle starts with a higher initial inclination, its orbit instead precesses about the eccentricity vector of the binary. In this configuration, the longitude of the ascending node of the particle orbit undergoes oscillations about $\phi_{\rm pb}=90\degree$ (or $\phi_{\rm pb}=270\degree$) while the tilt undergoes oscillations (this is the librating regime) \citep{Verrier2009,Farago2010,Doolin2011}.  The librating orbits are identified by the red curves while the green curves represent prograde circulating orbits and  the blue curves denote retrograde circulating orbits.  For a test particle ($j=0$), the critical inclination angle between the circulating and librating orbits is given by
\begin{equation}
i_{\rm crit} = \sin^{-1}\sqrt{\frac{1-e_{\rm b}^2}{1+4e_{\rm b}^2}}
\label{icrit0}
\end{equation}
\cite[e.g.,][]{Farago2010}.
For the eccentricity of 99 Her $(e_{\rm b} = 0.766)$, the critical angle is $i_{\rm crit} = 20.6\degree $, in agreement with the upper panel of Figure~\ref{fig:test_part}. 

We define the "stationary inclination", $i_{\rm s}$, as the inclination corresponding to the centre of the librating region. We show librating orbits 
that begin with $i<i_{\rm s}$ in red, and those that begin with $i>i_{\rm s}$ in cyan. For the test particle orbits, the binary orbit is fixed and so the orbit is the same no matter where it begins within the librating region. Thus, in this case the red and cyan lines are exactly the same. 

 The middle panel of Fig.~\ref{fig:test_part} shows particle orbits for $j = 0.1$ and the bottom panel shows particle orbits for $j = 0.51$. As the angular momentum ratio of the particle to the binary increases, the critical inclination  between prograde circulating and  librating orbits is higher. For $j >0$,  we apply  equations~31 and 37 of \cite{MartinLubow2019} to obtain for $j = 0.1$ that
 $i_{\rm crit} = 24.6^\circ$ and for $j = 0.51$ that $i_{\rm crit} = 40.7^\circ$. These values agree with the three-body simulations shown in the figure. 
 
For the higher mass particle, the binary orbit eccentricity oscillates and precesses. Consider a particle in a librating orbit that begins at $\phi_0=90^\circ$ with $i<i_{\rm s}$. When it reaches $\phi=90^\circ$ again, but with $i>i_{\rm s}$, the eccentricity of the binary has decreased. This means that the red and the cyan lines are no longer the same for a massive third body. We do not plot any circulating retrograde orbits for the high mass case because they are unstable \citep{Chen2019}.


While these three--body simulations are on stable repeating orbits, the presence of viscosity in a circumbinary accretion disc will act to damp its tilt oscillations.
Thus, the final inclination of the disc is either aligned (or counter aligned) to the binary orbit or polar aligned. Thus, the inclination at the centre of the librating region  is an important parameter for the disc simulations.  For the test particle case, $j=0$, the centre of the librating region corresponds to $i_{\rm s} = 90\degree$ and $\phi_{\rm s} = 90\degree$.  For the particle cases with $j=0.1$ and $j=0.51$, the centre of the librating regions corresponds to $i_{\rm s} = 87.9^\circ $ and $i_{\rm s} = 80.4^\circ $ \cite[calculated from equation 15 in][]{MartinLubow2019} and  $\phi_{\rm s} = 90\degree$, respectively, in agreement with Fig.~\ref{fig:test_part}.


\section{Circumbinary disc Simulations}
\label{sec:SPHsims}

The timescale for polar alignment may be shorter or longer than the lifetime of the gas disc, depending on the binary and disc parameters \citep{Martin2018}. Therefore, we explore the parameter space for the polar alignment of a circumbinary gas disc in 99 Her by varying the mass, inclination, and outer radius of the disc. The gas disc needs to evolve to a polar configuration within its lifetime for a polar debris disc to remain after the gas disc is dispersed.

\subsection{Simulation Setup}

We use the 3-dimensional smoothed particle hydrodynamics \cite[SPH; e.g.,][]{Price2012} code {\sc phantom} \citep{Lodato2010,Price2010,price2017}. {\sc phantom} has been well tested and used to model misaligned accretion discs in binary systems \citep[e.g.][]{Nixon2013,Martin2014,Franchini2019}. 
The disc simulations are in the so-called bending waves regime where the disc aspect ratio $H/R$ is larger than the  viscosity coefficient $\alpha$ \citep{shakura1973}. In this regime the warp induced in the disc by the binary torque propagates as a pressure wave with speed $c_{\rm s}/2$ \citep{Paploizou1983,papaloizou995b}. 

Each simulation consists of $N$ equal mass particles initially distributed from the inner disc radius, $R_{\rm in}$, to the outer disc radius, $R_{\rm out}$. The inner disc radius is chosen to be $R_{\rm in}=2a=33\, \rm au$, which is  close to the radius where tidal torque truncation is important \citep{Artymowicz1994}. However for a misaligned disc the tidal torque produced by the binary is much weaker, allowing the disc to survive closer to the binary \cite[e.g.,][]{Lubow2015,Miranda2015,Nixon2015,Lubow2018}.  For simulations with outer disc radius  $R_{\rm out} = 120\, \rm au$, we take $N=300,000$ particles,  and with $R_{\rm out} = 200\, \rm au$ we take $N=500,000$. The binary begins at apastron with an eccentricity of $0.766$. The accretion radius of each binary component is $R_{\rm acc}=4\, \rm au$. Particles within this radius are accreted and their mass and angular momentum are added to the star. We ignore the effect of self-gravity since it has no effect on the nodal precession rate of flat circumbinary discs.  For the narrow discs, the simulation lifetime is $1000\, \rm P_{orb}$ or $56,000\, \rm yr$, and for the extended discs it is $1500\, \rm P_{orb}$ or $90,000\, \rm yr$.

We chose to model the physical disc viscosity by using the artificial viscosity $\alpha^{\rm AV}$ , implemented in {\sc phantom} \citep{Lodato2010}.
The surface density profile of the disc is initially set as a power law distribution $\Sigma \propto R^{-3/2}$. The disc is locally isothermal with sound speed $c_s \propto R^{-3/4}$ and $H/R = 0.1$ at $R = R_{\rm in}$. With this prescription the disc is uniformly resolved meaning that $\langle h \rangle / H$ and therefore the disc viscosity parameter $\alpha$ are constant over the radial extent of the disc \citep{Lodato2007}. We take the \cite{shakura1973} $\alpha_{\rm SS}$ to be $0.01$. There exists a lower limit for the artificial viscosity in this type of simulation below which a physical viscosity is not resolved: $\alpha^{\rm AV} = 0.1$.

In our simulations, the disc is resolved with shell-averaged smoothing length per scale height $\langle h \rangle / H \approx 0.29$, independent of the disc size since we increase the number of particles for the larger disc radius. We show the initial parameters of the circumbinary disc for all simulations in Table~\ref{table2}.  We analyse the data from the SPH simulations by dividing the disc into 300 bins in spherical radius, $R$, which range from the radius of the inner-most particle to $170\, \rm au$ for narrow discs and $250\, \rm au$ for extended discs.
Within each bin we calculate the mean properties of the particles such as the surface density, inclination, longitude of ascending node, and eccentricity. The longitude of the ascending node of the disc is calculated by using the prescription from Equation (\ref{eq::phi_p}). 
In the next Section, we describe how varying the initial tilt, the initial disc mass, and the initial disc size affects the evolution.

\subsection{Alignment timescale calculations}

 In this section we describe the process for calculating the tilt decay timescale and for estimating the polar alignment timescale from simulations. As a disc evolves to a polar configuration, it undergoes librations of the longitude of the ascending node and tilt. Dissipation causes tilt oscillations that damp towards a polar configuration.  The tilt decay timescale, $\tau$, to the polar state  (that may correspond to an increase in tilt relative to the binary) is the time of exponential decay of the damped oscillations. The polar alignment timescale, $t_{\rm polar}$, is the time at which the disc is nearly fully orientated in a polar fashion. Several tilt decay timescales can be used as an estimate for the polar alignment timescale \citep{Martin2018}.


We determine the times of the first two local maxima of the inclination of the disc relative to the binary orbital plane to be $t_1$ and $t_2$. The tilt decay timescale towards the polar orientation is given by
\begin{equation}
    \tau = \frac{t_2 - t_1}{\log[(i_1-i_{\rm final})/(i_2-i_{\rm final})]}
    \label{tilt_decay}
\end{equation}
\citep{Martin2018}, where $i_1$ and $i_2$ are the local maxima of inclination at $t_1$ and $t_2$, respectively, and $i_{\rm final}$ is the final inclination of the disc. As can be seen in the following sections, some of the simulations exhibit strongly warped discs. Consequently, we a use the density-weighted average of the disc tilt. In all simulations, the resolution decreases with time. Thus, the tilt decay timescale in this manner needs to be calculated from early on in the simulations. 

Where possible, we define the polar alignment timescale, $t_{\rm polar}$, of the disc to be the time after which the magnitude of the difference in the density-weighted average of the longitude of the ascending node of the disc and the azimuthal angle of the binary eccentricity vector does not vary by more than $1^\circ$. Estimating the polar alignment timescale from  the inclination of the disc is difficult because the angular momentum of the disc evolves in time. The stationary inclination, $i_{\rm s}$,  increases as the disc loses mass \citep{MartinLubow2019}.
The tilt decay timescale and the polar alignment timescale from all simulation models are listed in the ninth and tenth columns of Table~\ref{table2}, respectively.

\begin{figure}
  \includegraphics[width=\columnwidth]{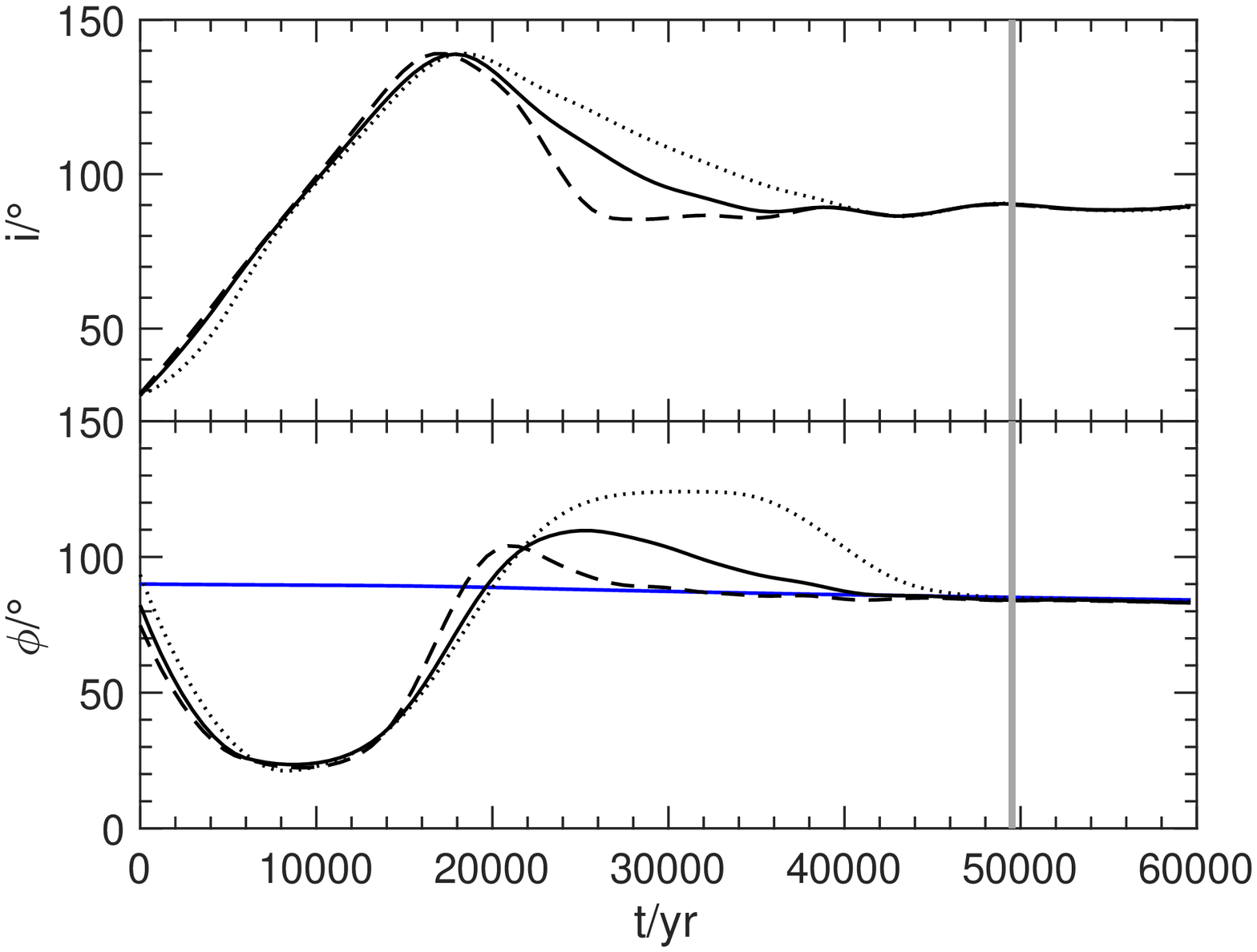}
  \includegraphics[width=\columnwidth]{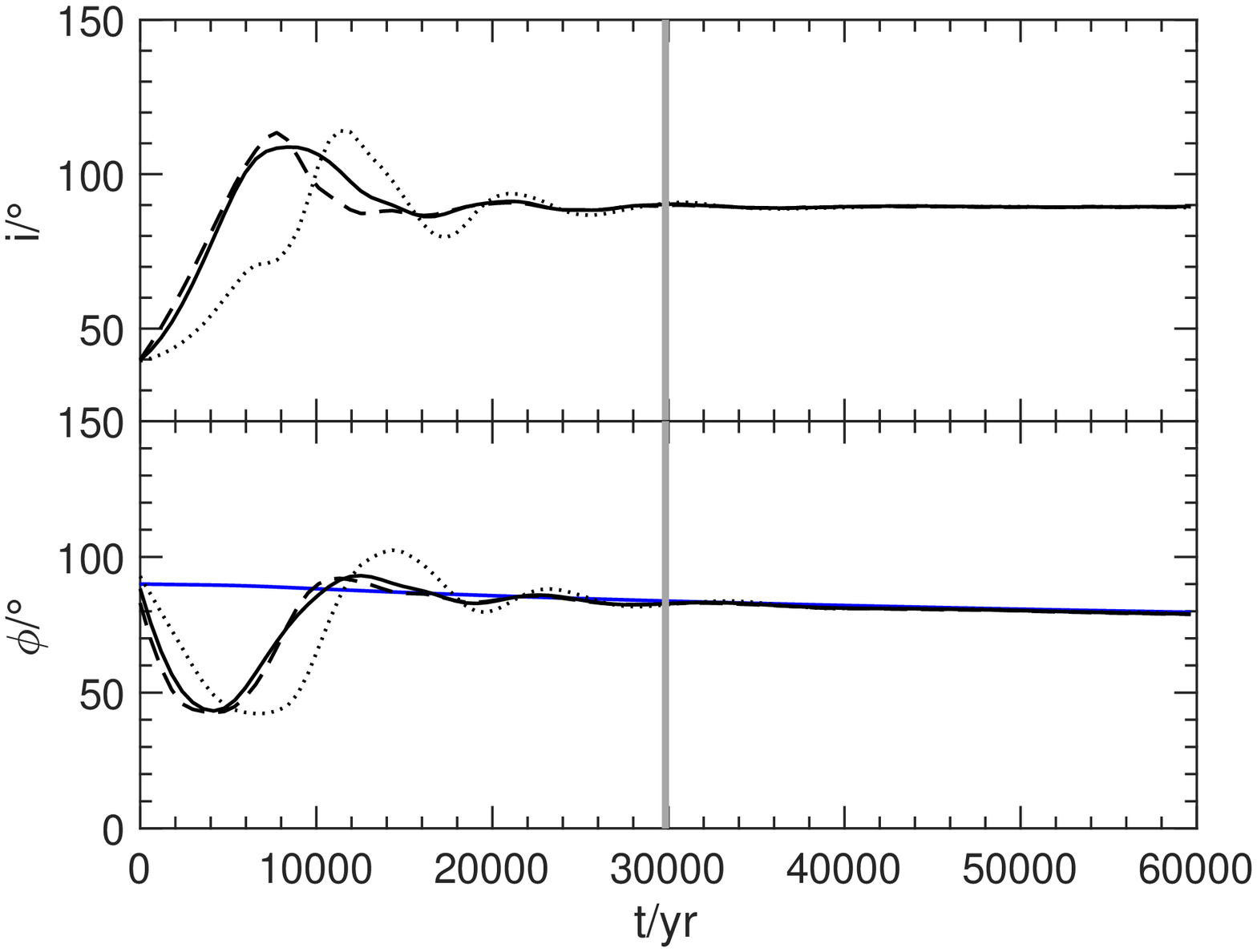}
  \includegraphics[width=\columnwidth]{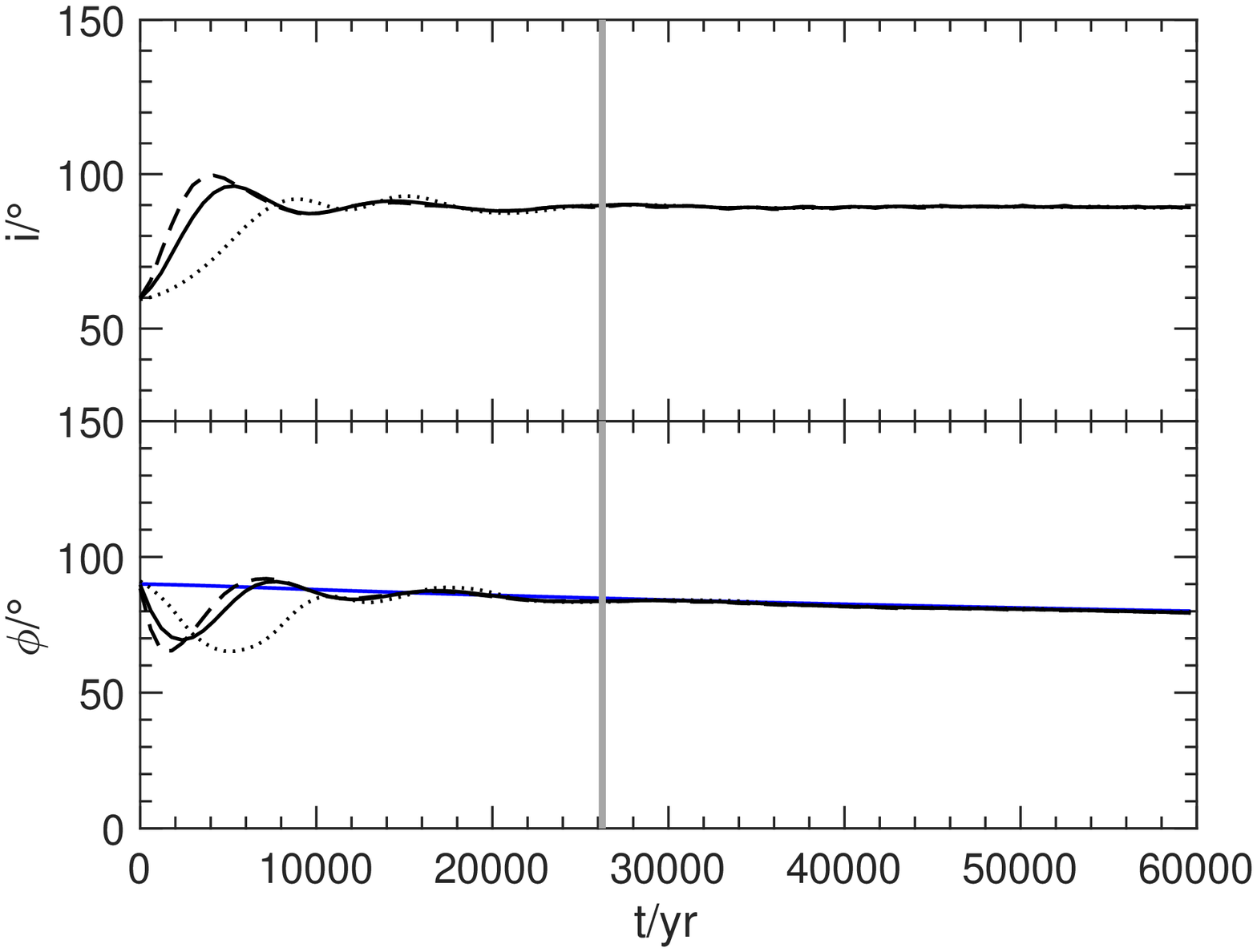}
  \caption{ Evolution of the inclination, $i$, and longitude of the ascending node, $\phi$, both as a function of time with varying initial disc tilt, $i_0$, evaluated at two radii within the disc, $50\, \rm au$ (dashed) and $120\, \rm au$ (dotted). The solid lines represent the density weighted  averages over the entire disc. The azimuthal angle of the eccentricity vector of the binary is shown by the blue line. The vertical gray line marks the polar alignment timescale.  Top panel: $i_0 = 30\degree $, (model B from Table~\ref{table2}). Middle panel: $i_0 = 40\degree $ (model C). Bottom panel: $i_0 = 60\degree $ (model D). }
\label{inc_plots}
\end{figure}

\begin{figure}
  \includegraphics[width=7.5cm]{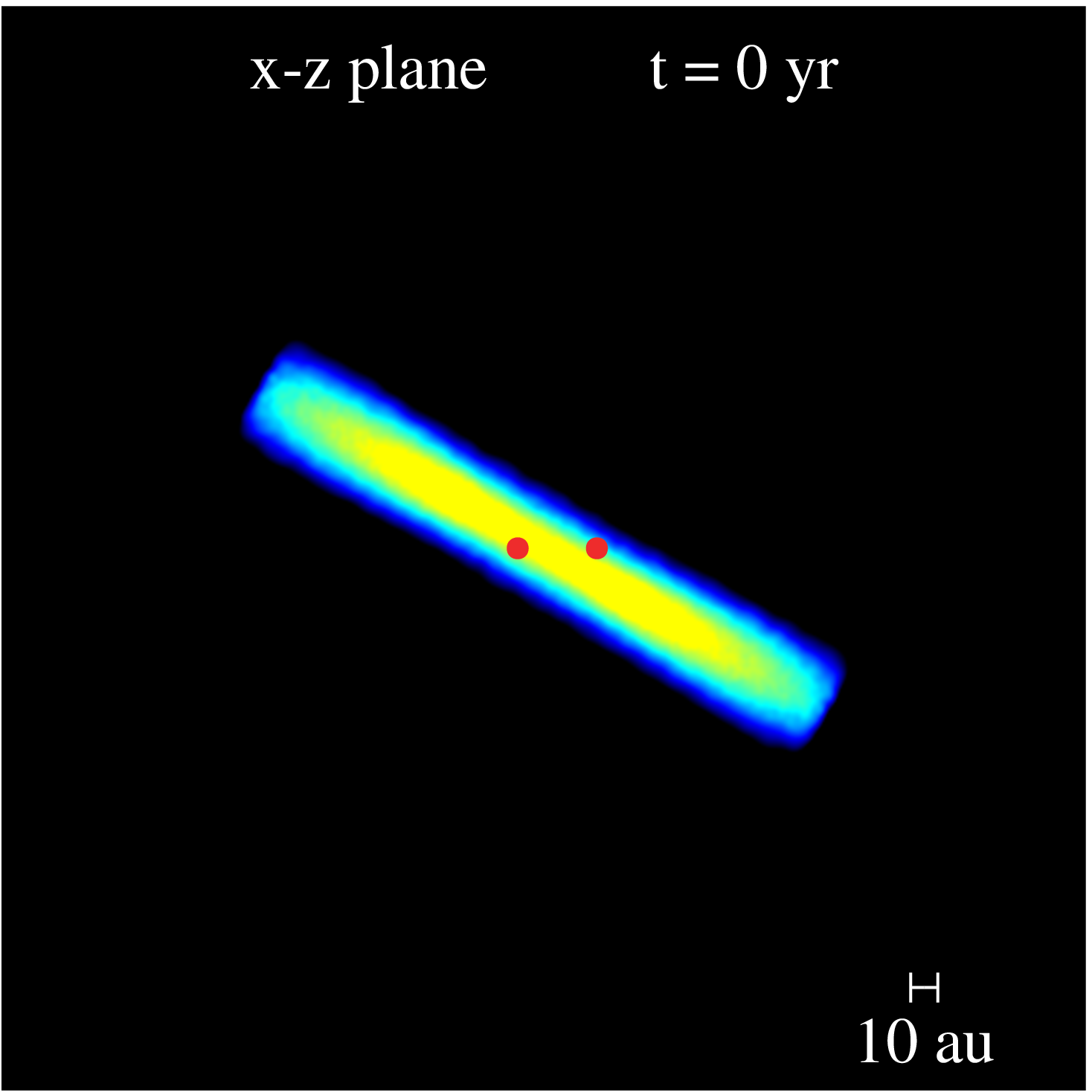}
  \includegraphics[width=7.5cm]{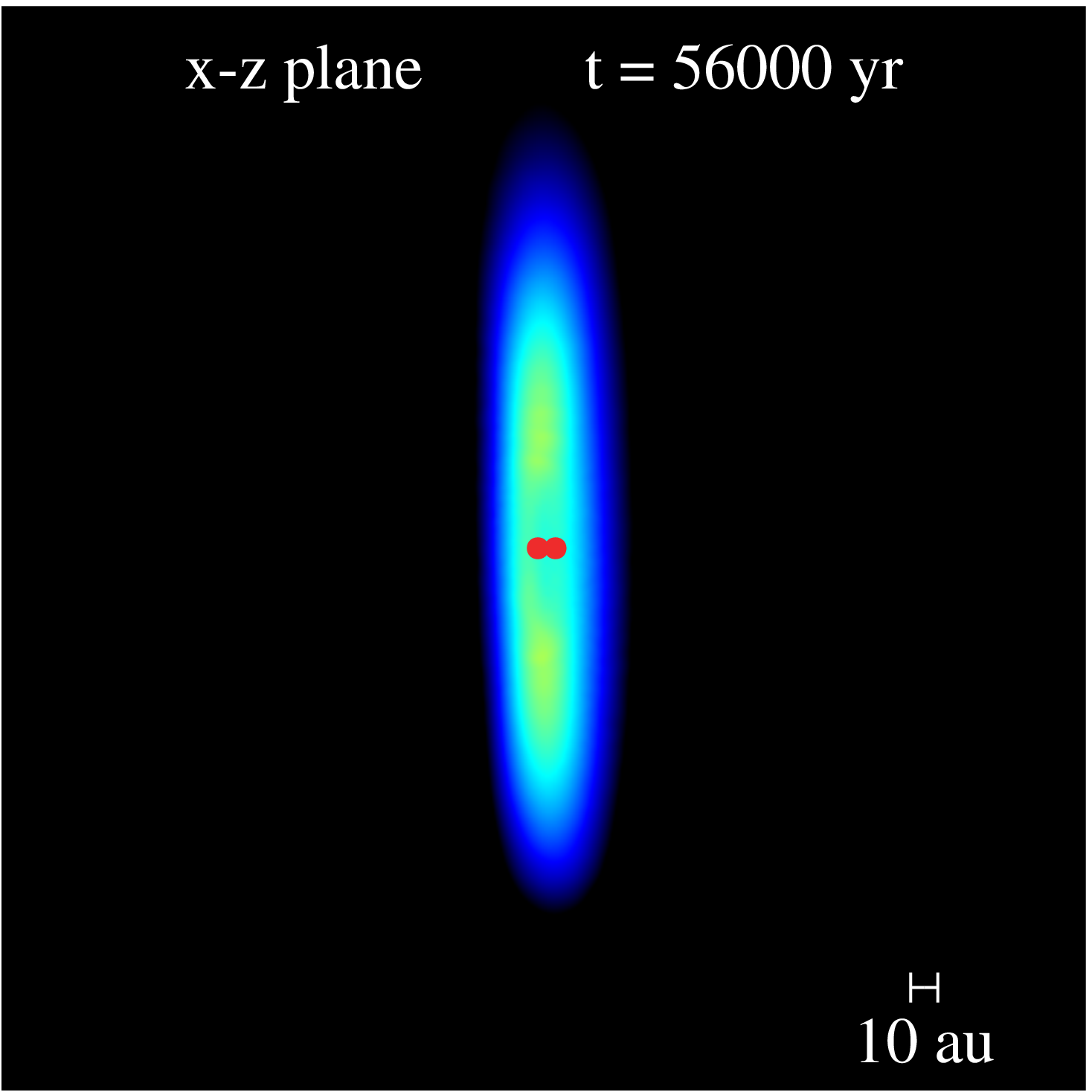}\centering
  \caption{Low mass circumbinary disc with an initial tilt of $30\degree$ (model B). The binary components are shown by the red circles, with the primary positioned to the left and the secondary to the right. The size of the circle is the accretion radius of the sink. Upper panel: initial disc setup for the SPH simulation in the $x$--$z$ plane. Lower panel: the disc in a polar configuration at a time of $t = 56,000\, \rm yr$. The color denotes the gas density with yellow regions being about two orders of magnitude larger than the blue.}
\label{splash}
\end{figure}




\subsection{Effect of initial disc tilt for a low mass disc}

We first vary the initial tilt of the disc while keeping the initial disc mass and initial disc size constant (models A, B, C and D in Table~\ref{table2}).  The initial outer radius of the disc is $R_{\rm out}=120\,\rm au$ and the initial disc mass is $0.001\,\rm M_\odot$.  Note that this initial value of $R_{\rm out}$ is the same as the ring radius obtained by \cite{Kennedy2012}. The disc outer radius increases as the disc evolves. Consequently, the disc could produce debris at  $120\,\rm au$. We consider the effects of larger initial disc outer radii in Section~\ref{sec:dsize}. The critical angle between librating and circulating solutions for a circumbinary test particle around 99 Her found in Section~\ref{sec:part_orbits} is $i=20.57\degree$. In these simulations the disc mass is very low ($M_{\rm d}=0.001\,M_{\odot}$) so the critical inclination angle that separates librating and circulating solutions is consistent with the test particle calculations.  The disc with initial misalignment of $20\degree$ (model A) aligns to the binary orbital plane rather than evolving towards a polar configuration. We do not show this result in the paper since we are interested in investigating discs that evolve to polar alignment.

Figure~\ref{inc_plots} shows the time evolution of the disc tilt, $i$, and the longitude of the ascending node, $\phi$, at two different radii, $50\, \rm au$ (dashed) and $120\, \rm au$ (dotted) inside the disc for the three simulations that go to polar alignment. We also show the density weighted average over the disc for both $i$ and $\phi$, given by the sold lines. The initial misalignment is $i_0=30\degree$ (model B), $40\degree$ (model C) and $60\degree$ (model D) for the top, middle and bottom panel respectively. The blue line represents the azimuthal angle of the eccentricity vector given by Equation (\ref{eq::phi_b}).

The tilt decay timescales, calculated from Equation (\ref{tilt_decay}), are $11,500$, $11,700$, and $17,000\, \rm yrs$ for discs with initial tilts of $i_0 = 30\degree$, $40\degree$ and $60\degree$, respectively. For all three of these models, the final inclination, $i_{\rm final}$, is $89\degree$. 
The tilt decay timescale decreases when the initial disc tilt becomes farther from the stationary inclination because the disc becomes more warped, leading to stronger dissipation and a faster decay timescale.
The polar alignment timescale for the three models are  $\sim 49,500\, \rm yr$,  $\sim 30,000\, \rm yr$, and $\sim 26,000\, \rm yr$, respectively. The estimated polar alignment timescales are a few tilt decay timescales. 
The polar alignment timescale is shown by the gray vertical lines in Fig.~\ref{inc_plots}. For the low mass disc, if its initial tilt is closer to the stationary inclination angle, then its evolution to polar alignment occurs on a shorter timescale. However, in all cases, the disc aligns to polar on a timescale much shorter than the expected disc lifetime.  

The evolution of a circumbinary disc whose initial tilt is close to the critical angle ($i=30^\circ$ and $i=40^\circ$) is somewhat different to the case with $i_0=60\degree$. A protoplanetary disc precesses nearly as a solid body if the radial communication timescale is shorter than the precession timescale \citep{papaloizou1995,Larwood1997}.
For a disc with a lower initial inclination, the librating region is larger which causes the disc to break into disjointed rings \citep[e.g.][]{Nixon2013}.
A disc with an initial tilt much greater than the critical inclination and closer to polar does not undergo this breaking and smoothly transitions to polar.

Figure~\ref{splash} shows the initial (upper panel) and final (lower panel) disc--binary system for the disc that is initially tilted by $30\degree$ (model B). The disc center initially lies at the center of mass of the binary and the binary angular momentum vector initially lies along the $z$--axis and the binary eccentricity vector initially along the $x$--axis. At a time of $56,000\, \rm yr$, the disc is in the polar configuration. 

While we have not run any simulations with inclination that is closer to retrograde than prograde, the alignment timescale of such a disc would be shorter than the prograde simulations shown here. A disc that is closer to retrograde feels a weaker binary torque and thus is able to extend closer to the binary \citep{Nixon2015,Miranda2015}. The precession timescale is shorter closer to the binary and therefore the alignment is faster \citep{Martin2018, Cuello2019}.

\begin{figure}
  \includegraphics[width=\columnwidth]{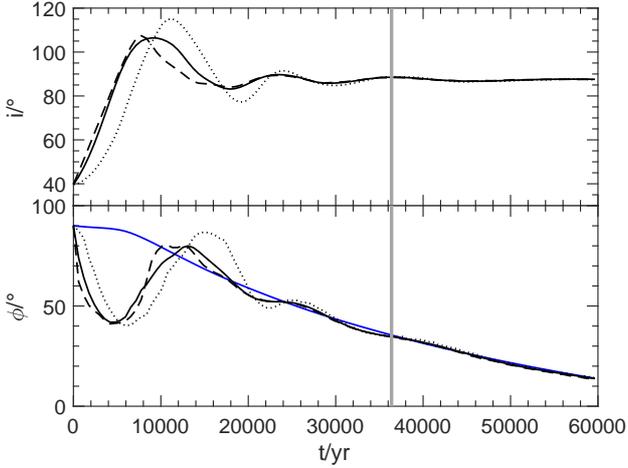}
  \caption{ Same as Fig.~\ref{inc_plots} but for a circumbinary disc with an initial tilt of $i_0=40\degree$ and initial disc mass of  $M_{\rm d}=0.01\, \rm M_{\odot}$ (model E).
  }
\label{mass_plots40}
\end{figure}

\begin{figure}
  \centering
  \includegraphics[width=\columnwidth]{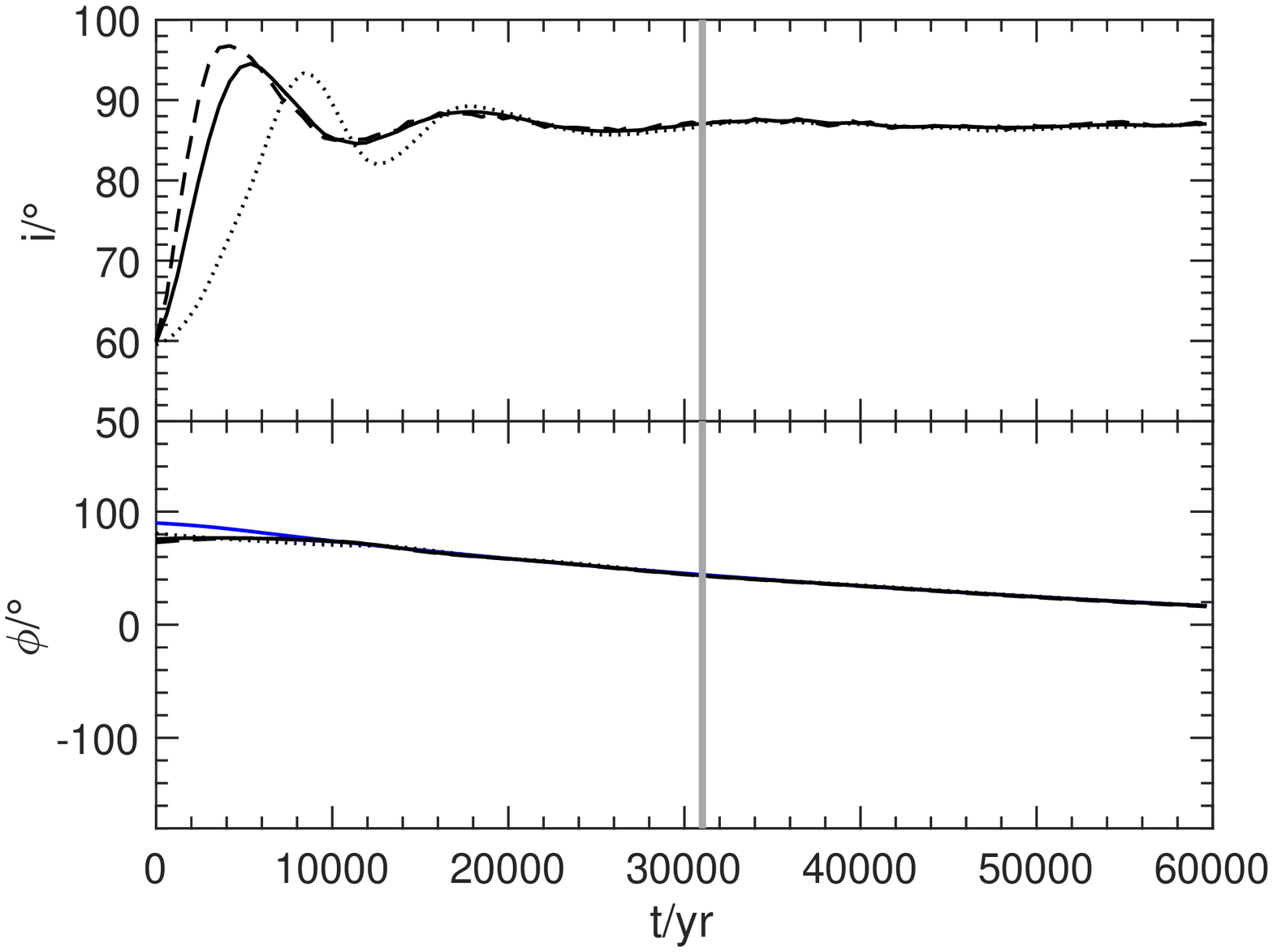}
  \includegraphics[width=\columnwidth]{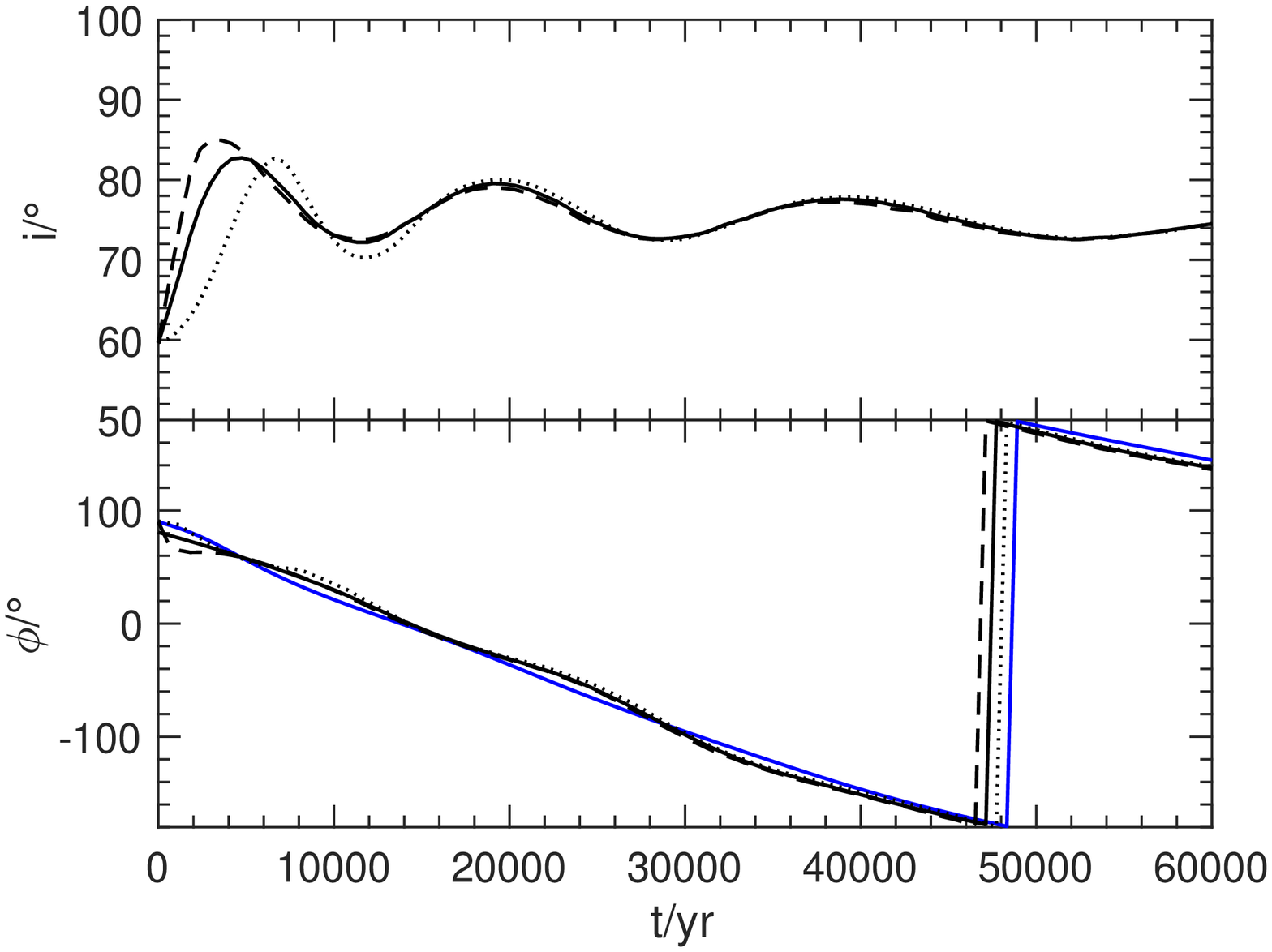}
  \caption{Same as Fig.~\ref{inc_plots} but with a circumbinary disc with $i_0=60\degree$. Top panel: $M_{\rm d}=0.01\, \rm M_{\odot}$ (model G). Bottom panel: $M_{\rm d}=0.05\, \rm M_{\odot}$ (model H). 
  }
\label{mass_plots60}
\end{figure}

\begin{figure}
    \centering
    \includegraphics[width =\columnwidth]{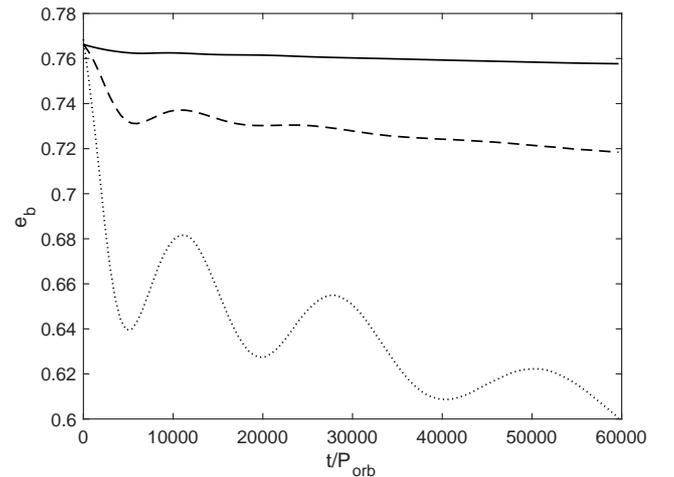}
    \caption{The time evolution of the binary eccentricity, $e_{\rm b}$, for three circumbinary disc masses: $M_{\rm d} = 0.001\, \rm M_{\odot}$ (solid), $M_{\rm d} = 0.01\, \rm M_{\odot}$ (dashed), and $M_{\rm d} = 0.05\, \rm M_{\odot}$ (dotted).}
    \label{fig:binary_e}
\end{figure}

\subsection{Effect of the disc mass}

We now additionally consider two higher values of initial disc mass, $M_{\rm d}=0.01\, \rm M_{\odot},\,0.05\, \rm M_{\odot}$ for initial tilts of $i_0=40\degree,\,60 \degree$. The ratio of the angular momentum of the disc to the angular momentum of the binary is initially $j=0.01$, $0.10,$ and $0.51$ for the initial disc masses $M_{\rm d}=0.001$, $0.01,$ and $0.05\,M$, respectively. 

Figure~\ref{mass_plots40} shows the evolution of the inclination and the longitude of the ascending node for an initial disc tilt of $i_0=40\degree$ with an initial disc mass $M_{\rm d}=0.01\, \rm M_{\odot}$ (model E),  which corresponds to $j=0.1$ initially. The critical inclination between librating and circulating solutions found from the three body problem in Section~3 is $i_{\rm crit} = 24.6^\circ$. Thus, this moderate mass disc evolves towards a polar configuration. Comparing to the low mass disc simulation with the same initial tilt (middle panel of Fig.~\ref{inc_plots}), the mass of the disc has not qualitatively changed the behaviour. The precession rate of the binary eccentricity vector is much higher for the higher mass disc as expected. However, the alignment timescale does not change significantly compared to the low mass disc.

We also considered a higher disc mass of $M_{\rm d}=0.05\, \rm M_{\odot}$ for a disc inclined by $40^\circ$ (model F). This disc has angular momentum ratio of the disc to the binary of $j = 0.51$ initially. In this case, the disc does not evolve to polar alignment but instead aligns to the binary orbital plane. For a particle with angular momentum ratio 0.51, we can estimate the critical inclination between librating and oscillating solutions to be $i_{\rm crit}=43.5^\circ$ \cite[with equation 37 in][]{MartinLubow2019}. Thus, the simulations agree with the three body problem solutions in Fig.~\ref{fig:test_part}. The higher the disc mass, the more likely it is that the disc will evolve to alignment with the binary orbital plane rather than polar alignment.


Figure~\ref{mass_plots60} shows the evolution of the inclination and longitude of the ascending node for different initial disc masses for a disc with initial inclination of $i_0=60\degree$. Even for high disc mass, providing that the initial inclination is sufficiently high, the disc still evolves to polar alignment. We estimate the stationary inclination angle for each simulation by taking the final inclination. The more massive the disc, the lower the stationary tilt inclination. This is consistent with our massive particle calculations in Section~\ref{sec:part_orbits} where we show that the center of the librating region, i.e. the stationary inclination angle, decreases as the mass of the particle increases. The stationary inclination angles inferred from our suite of  SPH simulations are $i_{\rm s}\approx 89\degree,\,87\degree,\,74.5\degree$ for discs with initial mass $M_{\rm d} = 0.001,\,0.01,\,0.05\, \rm M_{\odot}$ respectively. Since these disc masses correspond to $j=0.01$, $0.10,$ and $0.51$ initially, these correspond to the same angular momentum ratios as the particle orbits in Fig.~1.   In comparison, the particle model described in Section~\ref{sec:99Her} predicts  $i_{\rm s}\approx 90.0\degree, 87.9\degree, 80.4\degree$.  Consequently, the departures from polar alignment increase with mass in both the disc simulations and particle model cases. They are in rough quantitative agreement.



The discrepancy between the particle and disc results for the largest $j$ value is likely due in part to the effects of accretion torques on the binary that are not accounted for in the particle model.
The orbital evolution of the binary is dependent on both the accretion of angular momentum from the disc and the gravitational torques from the disc. Figure~\ref{fig:binary_e} shows the time evolution of the binary eccentricity for the three circumbinary disc masses. The most massive disc may eventually cause the binary to circularize. There is however, limited resolution in the inner gap which leads to uncertainties in the accretional torque (Martin \& Lubow 2019).   A lower binary eccentricity leads to a lower stationary inclination, as seen in the simulations. For a constant eccentricity, the stationary angle is shown as a function of the angular momentum ratio in the lower panel of Fig.~10 in \cite{MartinLubow2019} for a very similar eccentricity to that of 99 Her. Thus, in the following section, where we discuss the effects of disc size, we omit the high disc mass case due to the uncertain effects of accretional torques on the binary eccentricity evolution.


The models with an initial disc mass of $M_{\rm d}=0.01\, \rm M_{\odot}$ and initial tilts of $i_0 = 40\degree\,, 60\degree$, have tilt decay timescales of $15,000$ and $18,000\, \rm yrs$, respectively. The polar alignment timescale for the two models are estimated to be $36,000$ and $31,000\, \rm yr$, respectively.  As expected, a massive disc that has an initial tilt closer to the stationary inclination angle aligns  to that state on a shorter timescale.

The models with a disc mass of $M_{\rm d}=0.01\, \rm M_{\odot}$ and initial tilts of $i_0 = 40\degree\,, 60\degree$ undergo polar alignment within $\sim 23,000\, \rm yr$ and $\sim 12,000\, \rm yr$ respectively. A disc that has an initial tilt closer to the stationary inclination angle polar aligns on a faster timescale.  The disc with a mass of $0.05\, \rm M_{\odot}$ and an initial tilt of $60\degree$ has not fully aligned polar within the simulation time. Therefore, we can only estimate the tilt decay timescale, which is $67,000\, \rm yr$. The large tilt decay timescale implies that the polar alignment timescale for this massive disc would be of the order of $10^5\, \rm yr$. The comparison between discs with different masses ($M_{\rm d}=0.01\,, 0.05\, \rm M_{\odot}$) with the same initial inclination ($60\degree$) shows that a more massive disc aligns more slowly than a less massive disc. However, the alignment timescale is still shorter than the expected lifetime of a disc.

\begin{figure}
  \centering
  \includegraphics[width=\columnwidth]{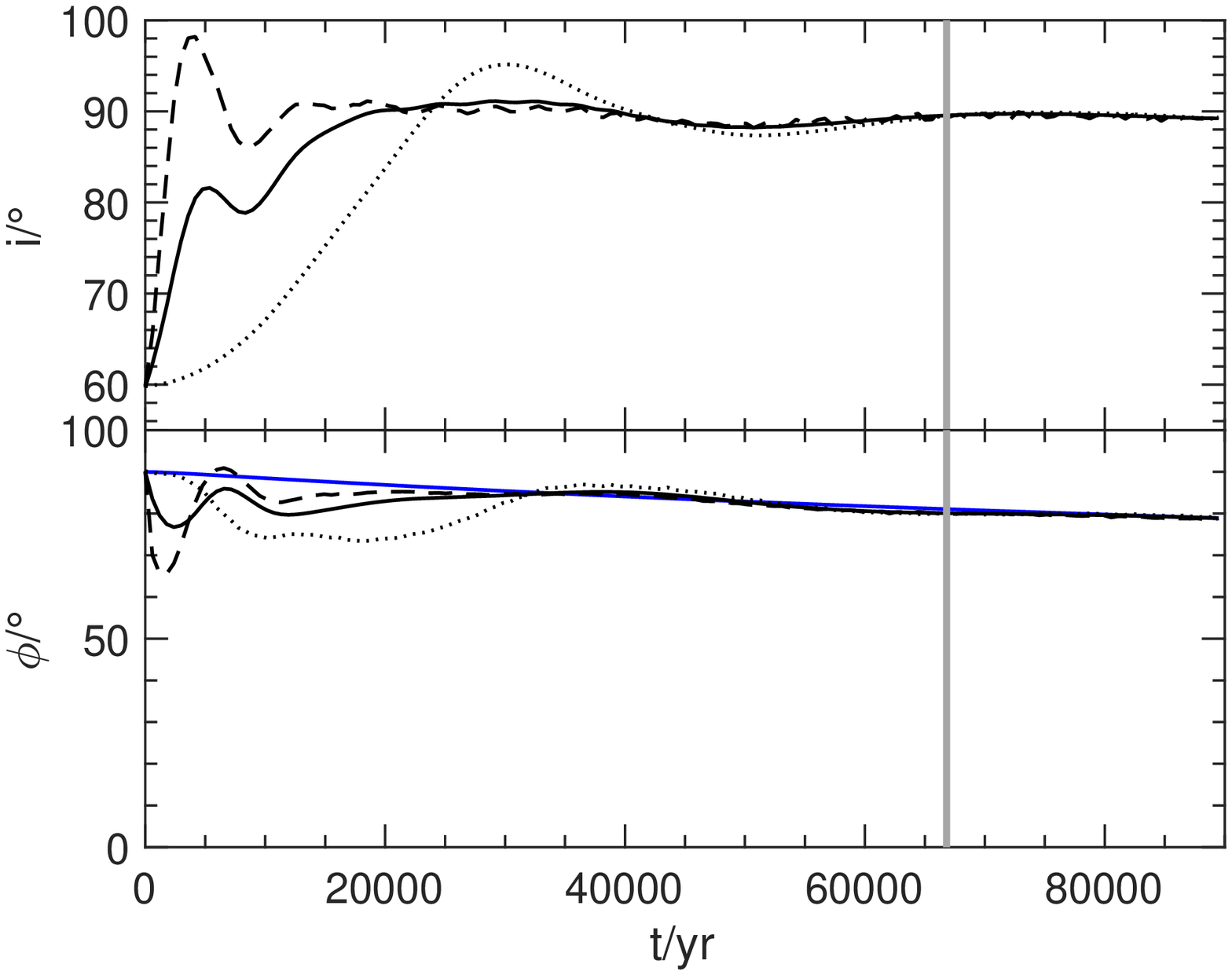}
  \includegraphics[width=\columnwidth]{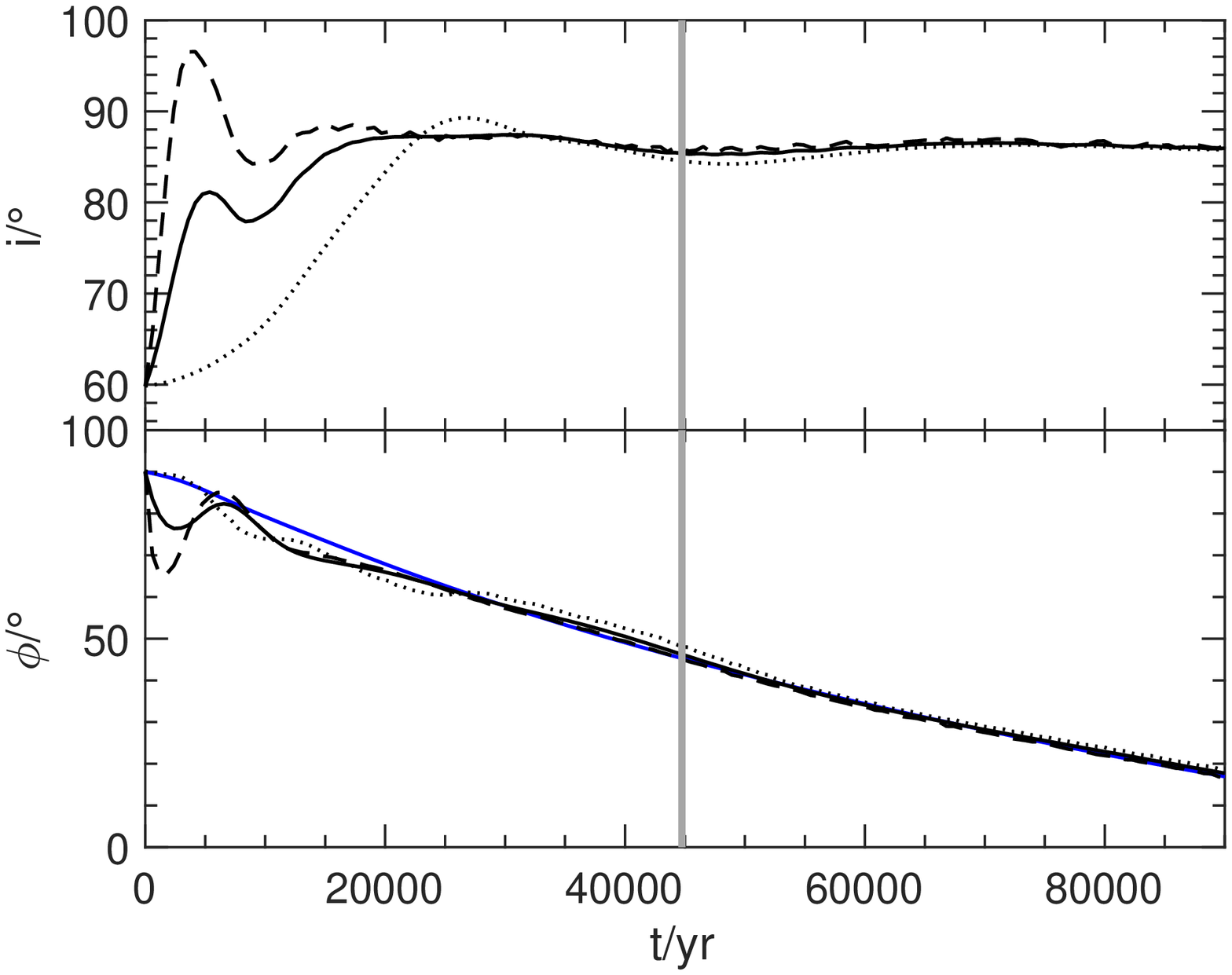}
  \caption{Evolution of the inclination, $i$, and longitude of the ascending node, $\phi$, both as a function of time for a circumbinary disc with an initial tilt of $i_0=60\degree$ and an outer radius of $r_{\rm out} = 200\, \rm au$. Measurements are evaluated at two radii within the disc, $50\, \rm au$ (dashed) and $200\, \rm au$ (dotted). The solid lines represent the density weighted averages over the entire disc. The azimuthal angle of the eccentricity vector of the binary is shown by the blue line. The vertical gray line marks the polar alignment timescale. Top panel: disc mass of $M_{\rm d}=0.001\, \rm M_{\odot}$ (model I).  Bottom panel: disc mass of $M_{\rm d}=0.01\, \rm M_{\odot}$ (model J).}
\label{fig::o200ab}
\end{figure}

\subsection{Effect of the disc size}
\label{sec:dsize}

Finally, we investigate how the disc size affects the polar alignment process in the context of the binary 99 Her. The models that have been described so far have investigated the behaviour of a disc that extends from $R_{\rm in}=33\, \rm au$ to $R_{\rm out}= 120\, \rm au$.
We now increase the disc outer radius to $R_{\rm out} = 200\, \rm au$ and increase the simulation time to $1500\, \rm P_{orb}$ or $~ 90,000\, \rm yr$. The initial disc tilt is set at $i_0=60\degree$. Because of uncertainties in the high disc mass model described in the previous subsection, we consider the two different lower initial disc masses,  $M_{\rm d}=0.001\, \rm M_{\odot}$ and $M_{\rm d}=0.01\, \rm M_{\odot}$ (see models I and J from Table~\ref{table2}). The initial ratio of the angular momentum of the disc to the initial angular momentum of the binary is $0.0122$ and $0.122$, respectively. We also increase the total initial number of equal mass particles to $N=500,000$ in order to uniformly resolve the disc as in the previous sections with a shell-averaged smoothing length per scale height of $\langle h \rangle / H \approx 0.29$.

Figure~\ref{fig::o200ab} shows the evolution of the tilt and the longitude of ascending node. The top panel represents the extended disc model with a mass of $M_{\rm d}=0.001\, \rm M_{\odot}$ and the bottom panel represents the model with a mass of $M_{\rm d}=0.01\, \rm M_{\odot}$. The inner regions of the disc undergo tilt oscillations on a shorter timescale compared to the outer regions, which slowly increase their tilt as the warp wave propagates outwards. A wider disc is less likely to precess as a rigid body because the radial communication timescale may be longer than the precession timescale, \cite[see section~4.5][for details]{Lubow2018}. Therefore the disc can become warped or it can even break \cite[e.g.,][]{NixonKing2012}. Figure \ref{fig::o200ab} shows a warp in the inclination and a twist in the precession angle that moves outwards in time for both disc masses.

Due to the intense warping in these larger discs, the tilt decay timescale cannot be calculated because it is no longer close to the linear regime. Though, the polar alignment timescale can still be estimated.  The less massive disc ($M_{\rm d}=0.001\, \rm M_{\odot}$) polar aligns after $t_{\rm polar} \sim 67,000\, \rm yr$, while the more massive disc ($M_{\rm d}=0.01\, \rm M_{\odot}$) has a polar alignment timescale of $t_{\rm polar} \sim 45,000\, \rm yr$. Therefore increasing the disc radial extent still results in a significantly shorter polar alignment timescale compared to the expected lifetime of a gas disc.

\begin{figure}
  \centering
  \includegraphics[width=\columnwidth]{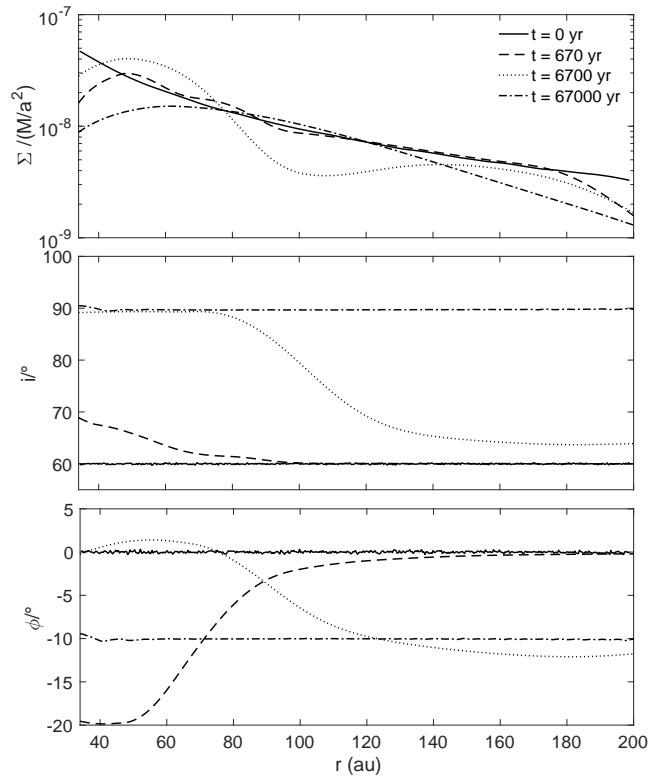}
  \caption{Surface density (top panel), tilt (middle panel) and longitude of the ascending node (bottom panel) as a function of radius at different times. The solid, dashed, dotted, and dashed--dotted curves  correspond to $t = 0, 6.7 \times 10^2, \, 6.7 \times 10^3, \, 6.7 \times 10^4 \, \rm yr$, respectively. The initial conditions of the circumbinary disc are for model I, which is an extended disc with a disc mass $M_{\rm d}=0.001\, \rm M_{\odot}$.}
\label{fig::params_run9}
\end{figure}




To investigate the local behaviour of the wider accretion disc around 99 Her, in Fig.~\ref{fig::params_run9} we show the surface density (top panel), inclination (middle panel), and longitude of the ascending node (bottom panel) as a function of radius at  $t=0\,,670\,, 6700\,,67,000\, \rm yr$ for a disc with initial mass $M_{\rm d}=0.001\, \rm M_{\odot}$. The initial surface density (at $t = 0\, \rm yr$, identified by the solid line) is a power law $\Sigma \propto r^{-3/2}$. As the disc evolves, the initial outer radius of the disc spreads outwards because of the presence of the disc viscosity. The inclination of the inner portions of the disc increase towards polar alignment as the wave travels outwards in time. 
Looking at the inclination of the disc as a function of radius at $6700\, \rm yr$ (dotted line) in the middle panel, we see that the disc inner regions inside about $100\, \rm au$ have larger misalignment compared to the outer regions. At the same time, the surface density profile shows a dip at around $100\, \rm au$.  The disc is strongly warped but  a higher resolution is required in order to properly resolve if the disc actually breaks \citep{Nixon2013}. Eventually, after roughly $t = 67,000\, \rm yr$ , the whole disc aligns polar as shown by the dot-dashed line in the middle panel of Fig.~\ref{fig::o200ab}.

\section{Analytic estimates}
\label{sec::linear_theory}

In this Section we first compare the alignment timescale calculated from linear theory to the results of the hydrodynamical simulations. We then use analytic results from the three body problem to constrain the mass of the gas disc at the time of disc dispersal based on the observed debris disc inclination. 

\subsection{Alignment timescale}

The main purpose of this work is to understand whether a primordial circumbinary accretion disc in 99 Her can undergo polar alignment within its lifetime, under a wide range of initial conditions, in order to explain the observations showing a polar debris disc around the binary star \citep{Kennedy2012}. We can see from the results of the simulations presented in the previous sections that the timescale on which the disc evolves to polar alignment is of the order of a few tens of thousands of years. The mechanism that leads to alignment between the accretion disc and the binary angular momenta is the natural presence of the disc viscosity that acts to damp the inclination oscillations on a timescale $\propto \alpha^{-1}$ dissipating the warp \citep{papaloizou1995,Lubow2000,Lubow2018}. The linear theory of warped discs describes the evolution of a disc that remains nearly flat and it has recently been applied to discs that are close to a polar aligned state \citep{Lubow2018,Zanazzi2018}.

The warp is dissipated and the disc aligns on a timescale
 \begin{equation}
 t_{\rm align} = \frac{1}{\alpha} \left(\frac{H}{R}\right)^2 \frac{\Omega_{\rm b}}{\Omega_{\rm p}^2}
 \label{eq:tbate}
 \end{equation}
 \citep[e.g.][]{Lubow2018},
 where $\Omega_{\rm b}=\sqrt{G(M_1+M_2)/a^3}$. For a disc that is aligning to polar, the  global disc precession frequency is given by
 \begin{equation}
     \Omega_{\rm p}=\frac{3\sqrt{5}}{4}e_{\rm b}\sqrt{1+4e_{\rm b}^2}\frac{M_1M_2}{M^2}\left<\left( \frac{a}{R}\right)^{7/2}\right>
 \end{equation}
 \citep[see equation 16 in][]{Lubow2018},
 where
 \begin{equation}
     \left<\left( \frac{a}{R}\right)^{7/2}\right>=\frac{\int_{R_{\rm in}}^{R_{\rm out}}\Sigma R^3 \Omega (a/R)^{7/2}\,dR }{\int_{R_{\rm in}}^{R_{\rm out}} \Sigma R^3 \Omega \, dr},
 \end{equation}
 where $\Omega=\sqrt{G(M_1+M_2)/R^3}$ is the Keplerian angular frequency. 
 Equation (\ref{eq:tbate}) is a rough estimate because it neglects the variation of quantities such as $\alpha$ and disc aspect ratio $H/R$ with radius $R$. Precise predictions of linear theory are made by computing the complex eigenfrequency for the linear tilt evolution equations \citep[e.g.,][]{Lubow2018}. But we do not carry out such calculations here.

\begin{figure}
    \centering
\includegraphics[width=\columnwidth]{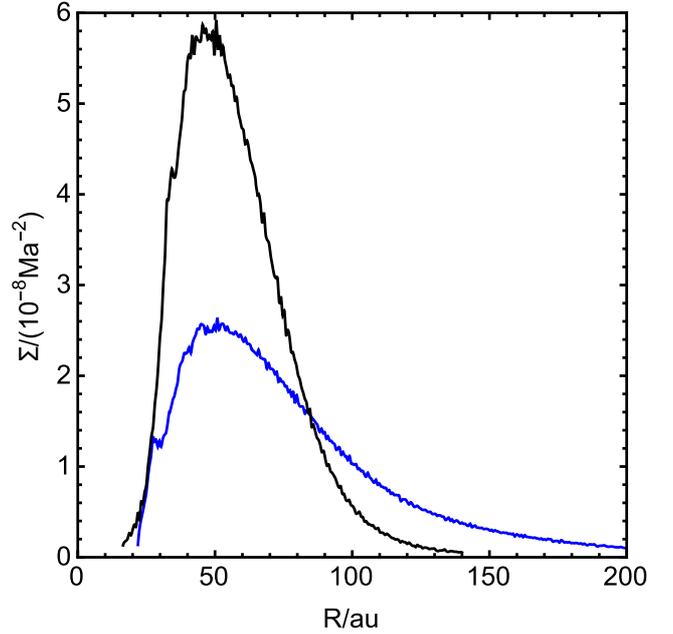}
    \caption{The surface density profile taken from model D at a time of $t=10,000\,\rm yr$ (black line) and from model I at a time of $t=30,000\,\rm yr$. 
    }
    \label{fig:align_sigma}
\end{figure}

\begin{figure}
    \centering
\includegraphics[width=\columnwidth]{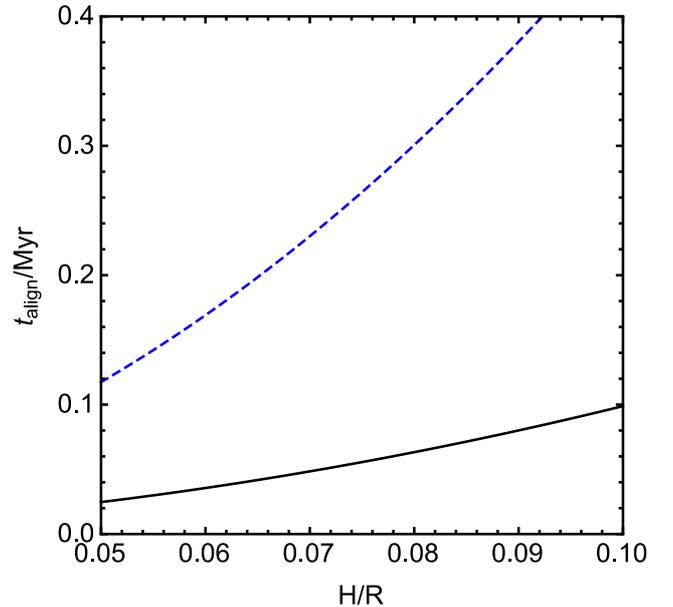}
    \caption{Alignment timescale calculated with equation~(\ref{eq:tbate}) as a function of $H/R$ for the surface density profiles taken from Fig.~\ref{fig:align_sigma}. The lower black solid line corresponds to the surface density profile taken from model D with initial truncation radius $R_{\rm out}=120\,\rm au$, while the blue dashed line corresponds to the surface density profile taken from model I with initial truncation radius $R_{\rm out}=200\,\rm au$.
    }
    \label{fig:align_sd}
\end{figure}

 The linear model we apply does not account for the evolution of the density distribution, as discussed in \cite{MartinLubow2019}. It takes as input the disc surface density profile that is approximated as being fixed in time. Since the density evolves over time, the profiles used in the calculation  are taken to be representative at some intermediate time.  We first take the surface density profile from the SPH simulations and calculate the alignment timescale. Fig.~\ref{fig:align_sigma} shows the surface density profile taken from model D at a time of $t=10,000\,\rm yr$ (black line) and from model I at a time of $t=30,000\,\rm yr$.  Fig.~\ref{fig:align_sd} shows the alignment timescale as a function of disc aspect ratio for the two surface density profiles with $\alpha=0.01$ and varying disc aspect ratio. 
The alignment timescale for the disc initially truncated at $120\,\rm au$ (model D with $H/R \approx 0.07$ at $r_{\rm out}$), $t_{\rm polar} = 26,000\,\rm yr$, is in rough agreement with the alignment timescale predicted by the black line in Fig.~\ref{fig:align_sd}, which is $\sim 50,000\, \rm yr$. For the larger initial truncation radius of $200\,\rm au$ (model I with $H/R \approx 0.06$ at $r_{\rm out}$), the alignment timescale predicted here, $\sim 140,000\, \rm yr$, is reasonable compared with  $t_{\rm polar} = 67,000\, \rm yr$ from the simulation. For both initial outer disc radii, the linear theory is off by a factor of two compared with simulations.


\begin{figure}
    \centering
\includegraphics[width=\columnwidth]{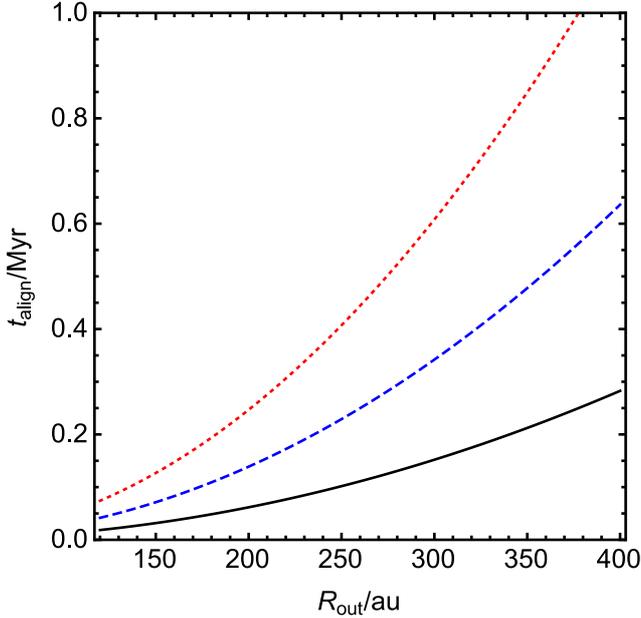}
    \caption{Alignment timescale calculated with equation~(\ref{eq:tbate}) as a function of the disc outer radius for a power law density profile $\Sigma \propto R^{-3/2}$, $R_{\rm in}=1.6\,a$ and $\alpha=0.01$. The solid black line has $H/R=0.05$, the dashed blue line has $H/R=0.075$ and the dotted red line has $H/R=0.1$.
    }
    \label{fig:align_th}
\end{figure}

Next, we consider a power law surface density with $\Sigma \propto R^{-3/2}$ extending from $R_{\rm in}=1.6\,a$ \citep[see][for a discussion of the inner disc radius of a polar aligned disc]{Franchini2019b} up to $R_{\rm out}$. We take $\alpha=0.01$ and consider different values for $H/R$.
The results are shown in Figure \ref{fig:align_th} as a function of the disc outer radius.  A radially wider accretion disc is expected to align on a longer timescale compared to a more narrow disc with the same mass and initial inclination.  This is consistent with the comparison between the results for model D (shown in the bottom panel of Figure \ref{inc_plots}) and model I, shown in the upper panel of Figure \ref{fig::o200ab}. Even if the gas disc is extended significantly father out than our simulations, the alignment timescale is still shorter than the expected disc lifetime.  However, this holds for $\alpha=0.01$ used in the simulations. For much smaller $\alpha \la 10^{-4}$, Equation (\ref{eq:tbate}) predicts that the polar alignment timescale could become longer than the disc lifetime. 

Dust grains are not included in our simulations but the time at which they decouple from the disc provides a more stringent constraint on the required polar alignment timescale than the lifetime of the disc.  The dust particles must be coupled to the gas disc until it reaches polar alignment. Otherwise, a misaligned disc of debris will undergo differentiated precession and lead to spherical distribution rather than a disc \citep[e.g.][]{Nesvold2016}. Therefore, the time it takes the dust particles to grow and decouple from the gas must be longer than the polar alignment timescale of the gas disc.

\begin{figure}
    \centering
\includegraphics[width=\columnwidth]{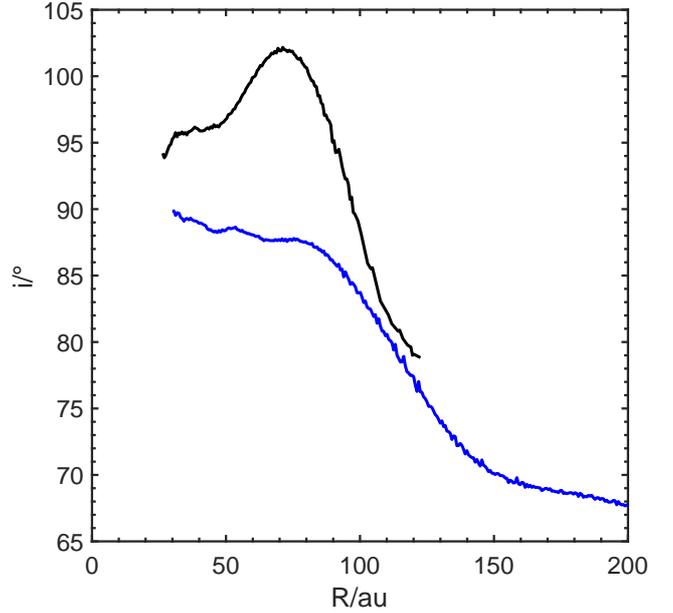}
    \caption{The inclination profile taken from model D at a time of $t=5,000\,\rm yr$ (black line) and from model I at a time of $t=10,000\,\rm yr$ (blue line).}
    \label{fig:warp_fig}
\end{figure}

The strong disc warping can be seen in Fig.~\ref{fig:warp_fig}, where  the disc tilt is given as a function of radius with $<h>/H \leq 0.5$. the black line shows the narrow disc (model D) at a time of $t = 5,000\, \rm yr$ and the blue shows the wider disc (model I) for a time of $t = 10,000\, \rm yr$.    Strong nonlinear disc warping is beyond the regime of applicability for linear theory. Strong warping may lead to additional dissipation and reduce the alignment timescale to below the values we estimate.


We see from Figs.~\ref{inc_plots} and~\ref{mass_plots60} that less massive discs tend to align polar on a slightly shorter timescale, regardless of the initial misalignment while more massive discs tend to align on a longer timescale. The linear theory we apply assumes that the disc mass/angular momentum is very small compared to the binary mass/angular momentum. However, in general we find the theoretical estimates to be consistent with the timescale for the disc to evolve to polar alignment inferred from our SPH simulations.


\subsection{Stationary inclination}

Since the timescale for polar alignment is shorter than the disc lifetime for  a broad range of disc parameters, the debris disc should be at an inclination given by the generalised polar state. In the limit of a zero angular momentum disc, the generalised polar state is a disc aligned to the binary eccentricity vector and inclined to the binary angular momentum vector by $90^\circ$. A more massive disc has a generalised polar alignment with a lower level of misalignment. 

We follow \cite{MartinLubow2019} and calculate the maximum possible disc gas mass at the time of decoupling of gas and solids given that the observed debris disc is within $3^\circ$ of being polar aligned.  The decoupling may occur when the gas disc disperses. We assume that the disc has aligned to a stationary inclination that depends upon binary eccentricity and disc angular momentum given by
\begin{eqnarray}
  \cos i_{\rm s} =  \frac{ -(1+ 4e_{\rm b}^2)+ \sqrt{ \left(1+ 4e_{\rm b}^2 \right)^2+60 (1-e_{\rm b}^2) j^2} }{10 j}
    \label{farz}
\end{eqnarray}
\citep[see equation 17 in][]{MartinLubow2019}. This was derived using the secular theory of a circumbinary particle with mass \citep{Farago2010}.  Since we know the binary eccentricity and we set $i_{\rm s}=87^\circ$, we can find an upper limit on the angular momentum ratio of 
\begin{equation}
j=\frac{J_{\rm d}}{J_{\rm b}}=0.14,
\label{eqj}
\end{equation}
where $J_{\rm d}$ is the angular momentum of the disc.
We assume that the surface density of the disc is distributed as a power $\Sigma \propto R^{-q}$ between $R_{\rm in}=1.6\,a=26.4\,\rm au$  and $R_{\rm out}$.
We parameterise the disc angular momentum with
\begin{equation}
 J_{\rm d} = k M_{\rm d} \, a^2 \Omega_{\rm b},
 \label{jvn1}
 \end{equation}
 where $k$ depends upon the density profile. For $R_{\rm out}=120\,\rm au$, $k$ varies from 2.2 to 2.0 as $q$ varies from 0 to 1.5. For $R_{\rm out}=200\,\rm au$, $k$ varies from 2.8 to 2.4 as $q$ varies from 0 to 1.5. 
The angular momentum of the binary is
\begin{equation}
J_{\rm b}= 0.14 M a_{\rm b}^2 \Omega_{\rm b}.
\label{jbn}
\end{equation}
For $R_{\rm out}=120\,\rm au$ and $q=1.5$, with equation~(\ref{eqj}) we find the disc mass to be $M_{\rm d} \approx 0.01 M=0.014\,\rm M_\odot$. 
 But since the 99 Her debris ring has a radius of about $120\,\rm au$, we must consider larger discs.
For $R_{\rm out}=200\,\rm au$ and $q=1.5$  we find the disc mass to be $M_{\rm d} \approx 0.0085 M=0.012\,\rm M_\odot$.  If the debris disc has an inclination of less than $3^\circ$ to polar, the disc mass may be lower than these estimates. These masses are in agreement with  SPH simulations. Models E, and G that begin with a disc mass of $0.01\,M$ both have a final inclination of $87\degree$. This disc mass is also quite reasonable for a protostellar disc. This mass represents the gas disc mass at the time that the gas and the dust-producing solids within the disc decouple. In principle, this imposes further constraints that we do not examine here. 
\section{Discussion}
\label{sec:dis}

Hundreds of debris discs around single stars have been detected to date in far-IR surveys. Since the dust has a lifetime shorter than the age of the host star these observations imply that the dust contained in the debris disc is likely not primordial, at least for main sequence stars with age $> 10\,{\rm Myr}$ \citep{Wyatt2018}. The dust content is therefore thought to be replenished through planetesimal collisions. 
Therefore, in principle, debris discs provide information on the planet formation process showing the location and properties of the planetesimals and also their collision velocities.
 
The detected extrasolar debris discs have shown a variety of morphologies \citep{Booth2013}, including evidence of multiple components similar to the outer Solar system configuration \citep{Chen2014,Kennedy2014}. This provides only circumstantial support to the idea of the presence of planets within debris disc systems since they can remove dust from certain regions of the disc.

All currently known imaged planets have been observed in systems with debris discs \citep{Bowler2016,Ricci2015,Marino2016}. In principle, it is then possible to study imaged planets and discs in the same system while this is difficult in protoplanetary discs due to their high optical depths \citep{Hughes2018}. Several efforts are currently being made to find a possible correlation between debris discs properties and imaged planets. Since the planets and the debris disc typical locations are separated by tens of au, it is not clear how the planets influence the disc detectability. 

There are currently 21 confirmed circumbinary planets, of which 10 discovered through transit within 9 binary systems \citep{Doyle2011,Welsh2012,Orosz2012a,Orosz2012b,Schamb2013,Kostov2013,Kostov2014,Welsh2015,Kostov2016,Orosz2019}. For these systems, the planet orbital inclination with respect to the binary orbital plane is very low \citep{Kostov2014,Welsh2015}.
This is likely the result of the small orbital period of all the observed binaries leading to tidal circularisation of the binary and therefore alignment of the disc \citep{Czekala2019}. We showed that a sufficiently misaligned and not too massive accretion disc around an eccentric binary can undergo polar alignment within its lifetime  and therefore form a polar debris disc.
This implies that if planets formed in this disc they would be on polar orbits around the binary. However, there are no confirmed circumpolar planets detected so far. Planets with polar orbits would be harder to detect than the nearly coplanar planets found by recurrent transits with Kepler. Polar planets may be detectable as nonrecurring  transits of the binary or eclipse timing variations of the binary \citep{ZhangFabrycky2019}.  The best fit model to reproduce the 99 Her observations is a polar debris ring \citep{Kennedy2012}. A polar ring could be produced by the presence of polar planets. It is known that shepherding planets can cause a debris disc to form ring-like structures \citep{Rodigas2014}. In addition, the debris disc may contain or evolve to contain polar planets. 

\section{Conclusions}
\label{sec:conc}

We have explored parameters of a circumbinary gas disc  around the eccentric binary system 99 Her that result in polar alignment of the disc.
We investigated the effects of the initial disc inclination, mass and size on the timescale over which the disc aligns polar to the binary orbital plane. Since the eccentricity of 99 Her is very high ($e_{\rm b}=0.766$), the initial disc misalignment does not have to be very large for the disc to evolve to polar alignment, as suggested by \cite{Martinlubow2017}.
The critical angle depends on the mass of the disc and is larger for a more massive disc (see Fig.~1).

Since the observed inclination of the debris disc relative to the binary is only a few degrees away from $90\degree$, the initial mass of the gas disc can be constrained to be $M_{\rm di} \lesssim 0.01$. For low mass discs, an initially mild inclination ($i_0 > 20\degree$) results in polar alignment, a more massive disc is able to align polar only if the initial misalignment is moderate (e.g., $i_0 > 40\degree$).
Gas discs that are initially inclined by more than $60\degree$ undergo polar alignment regardless of the initial disc mass.
Wider gas discs are still able to align polar even though they might warp and break if the sound crossing timescale is longer than the precession timescale induced by the binary torque.

All simulations that evolved to a polar state, did so on timescales of order of tens of thousands of years. The polar alignment timescale for these simulations is therefore much shorter than the lifetime of the gas disc, which has been estimated  for circumstellar gas discs to be a few Myr \citep{Haisch2001}  or longer for circumbinary gas discs, as discussed in the Introduction.  We note that the time at which the dust grains decouple from the gas disc may provide a more stringent constraint than the  lifetime of the gas disc. 
Moreover, our simulations have a turbulent viscosity parameter $\alpha=0.01$, while the linear theory suggests that for significantly smaller disc viscosity $\alpha \la 10^{-4}$ the alignment timescale becomes longer than the disc lifetime. 
 Mindful of these caveats, we conclude that the presence of a remnant polar debris disc around 99 Her can be explained by an initially inclined gas disc evolving into a polar configuration before dispersing the gas under a wide range of initial conditions and disc parameters. Such a disc may provide an environment for the formation of polar planets.

\section*{Acknowledgements}

 We thank the anonymous referee for helpful suggestions that positively impacted the work.  We thank Daniel Price for providing the {\sc phantom} code for SPH simulations and acknowledge the use of {\sc splash} \citep{Price2007} for the rendering of the figures.   Computer support was provided by UNLV's National Supercomputing Center. We acknowledge support from NASA through grants NNX17AB96G and 80NSSC19K0443.







\bibliographystyle{mnras}
\bibliography{ref} 

\bsp	
\label{lastpage}
\end{document}